\documentstyle[epsfig]{mn}
\newif\ifAMStwofonts
\newcommand{\be}{\begin{equation}}
\newcommand{\ee}{\end{equation}}
\newcommand{\bea}{\begin{eqnarray}}
\newcommand{\eea}{\end{eqnarray}}
\newcommand{\nn}{\nonumber}

\newcommand{\noi}{\noindent}
\newcommand{\appgeq}{\stackrel{>}{\sim}}
\newcommand{\appleq}{\stackrel{<}{\sim}}

\newcommand{\di}{\partial}
\newcommand{\Bphi}{B_\phi}
\newcommand{\Bz}{B_z}

\newcommand{\Bphis}{B_{\phi S}}
\newcommand{\Bzs}{B_{z S}}
\newcommand{\Ps}{P_S}      
\newcommand{\Pmag}{P_{mag}}
\newcommand{\Pmags}{P_{mag,S}}
\newcommand{\Rs}{{R_S}}

\newcommand{\sigsq}{{\sigma^2}}

\newcommand{\Pave}{\langle P \rangle}
\newcommand{\Pmagave}{\langle P_{mag} \rangle}

\newcommand{\rhoave}{\langle \rho \rangle}

\newcommand{\sigsqave}{\langle \sigma^2 \rangle}
\newcommand{\Alfven}{Alfv\'en }
\newcommand{\Alfvenic}{Alfv\'enic }   
\newcommand{\Stod}{Stod\'olkiewicz }

\newcommand{\Gz}{\Gamma_z}
\newcommand{\Gphi}{\Gamma_\phi}
\newcommand{\amag}{\alpha_{mag}}
\newcommand{\anon}{\alpha_{non}}
\newcommand{\W}{{\cal W}}
\newcommand{\M}{{\cal M}}
\newcommand{\V}{{\cal V}} 
\newcommand{\dr}{\frac{d}{dr}}
\newcommand{\ds}{\frac{d}{ds}}
\newcommand{\Pm}{P_{mag}}
\newcommand{\lowbphi}{b_\phi}
\newcommand{\fz}{f_z}
\newcommand{\fphi}{f_\phi}
\newcommand{\fpg}{4\pi G}

\newcommand{\mh}{m_h}
\newcommand{\mvir}{m_{vir}}
\newcommand{\mmag}{m_{mag}}

\newcommand{\thirteenCO}{^{13}CO}
\newcommand{\twelveCO}{^{12}CO}

\newcommand{\Msol}{M_\odot}
\newcommand{\x}{\times}

\title{Helical Fields and Filamentary Molecular Clouds}
\author[Jason D. Fiege and Ralph E. Pudritz]
       {Jason D. Fiege and Ralph E. Pudritz \\
	Dept. of Physics and Astronomy \\
	McMaster University \\
	1280 Main St. W.,
	Hamilton, Ontario \\
	L8S 4M1 \\
	email: fiege@physics.mcmaster.ca, \\
	pudritz@physics.mcmaster.ca}
\date{\today}  
      
\pagerange{\pageref{firstpage}--\pageref{lastpage}}
\pubyear{1998}
       
\begin{document}

\maketitle

\label{firstpage}
 
\begin{abstract}

\bigskip

We study the equilibrium of pressure truncated, filamentary molecular clouds 
that are threaded by rather general helical magnetic fields.
We first derive a new form of the virial equation
appropriate for magnetized filamentary clouds, which includes the effects of 
non-thermal motions and the turbulent pressure of the surrounding ISM.
When compared with the data, 
we find that many filamentary clouds have a mass per unit length that is
significantly reduced by the effects of external pressure,
and that toroidal fields play a significant role in squeezing such clouds.

We also develop exact numerical MHD models of 
filamentary molecular clouds with more general helical field configurations than have previously been
considered.  We examine the effects of the
equation of state by comparing ``isothermal'' filaments, with constant
total (thermal plus turbulent) velocity dispersion, with equilibria constructed using a logatropic
equation of state.

Our theoretical models involve 3 parameters; two to describe the mass loading
of the toroidal and poloidal fields, and a third that describes the radial concentration of the 
filament.  We perform a Monte Carlo exploration of our parameter space to determine which 
choices of parameters result in models that agree with the available observational constraints.
We find that both equations of state result in equilibria that agree with the observational
results.  Moreover, we find that models with helical fields have more realistic density profiles than 
either unmagnetized models or those with purely poloidal fields; we find that most isothermal models 
have density distributions that fall off as $r^{-1.8}$ to $r^{-2}$, while logatropes have density profiles that
range from $r^{-1}$ to $r^{-1.8}$.  
We find that purely poloidal fields produce filaments with steep radial density gradients
that are not allowed by the observations.

\end{abstract}
      
       
\section{Introduction}
\renewcommand{\textfraction}{0} 
\renewcommand{\topfraction}{1} 
\renewcommand{\bottomfraction}{1} 

Observations have revealed that most molecular clouds are filamentary structures that are
supported by non-thermal, small-scale MHD motions of some kind, as
well as large scale ordered magnetic fields (cf. Schleuning 1998).
Nevertheless, virtually all theoretical models assume
spheroidal geometry.  
While spheroidal models are a reasonable geometry for molecular cloud cores, 
these cannot adequately describe molecular clouds on larger scales.
The goal of this paper is to fully develop a theory for filamentary molecular clouds
including the effects of ordered magnetic fields.  
It is our intent that this work should elevate
filamentary clouds to the same level of understanding as that enjoyed by their
spheroidal counterparts (cf. review McKee et al. 1993).
This is an important step in star formation theory because filamentary molecular
clouds ultimately provide the initial conditions for star formation.  
A clear understanding of the initial conditions is necessary if we are to understand 
the processes by which clouds produce their star-forming cores.

Ostriker (1964) investigated the equilibrium of unmagnetized isothermal filaments;
he found that the density varies as $\sim r^{-4}$ in the outer regions.   However, this solution is
much too steep to account for the observed density profiles in molecular clouds.
For example, Alves et al. (1998, hereafter A98) and Lada, Alves, and Lada (1998, hereafter LAL98) use extinction measurements of
background starlight in the near infra-red to find $r^{-2}$ density
profiles for the filamentary clouds L977 and IC 5146.

Most theoretical models for self-gravitating filaments have featured magnetic fields
that are aligned with the major axis of the filaments.  The pioneering work by Chandrasekhar and Fermi (1953) 
was the first to analyse the stability of magnetized incompressible
filaments with longitudinal magnetic fields.  
\Stod (1963) developed a class of isothermal models in which the ratio of the gas to magnetic pressure 
($\beta$) is constant.  The magnetic field in 
these models simply re-scales the Ostriker (1964) solution; thus, the steep $r^{-4}$ density profile is preserved.
The structure of the Ostriker solution is unchanged by the addition of a uniform poloidal magnetic field.
The stability of such models, including the effects of pressure truncation, has subsequently been determined by
Nagasawa (1987).  Gehman, Adams, \& Watkins (1996) have considered the effects of a logatropic equation of state (EOS)
on the equilibrium and stability of filamentary clouds threaded by a uniform poloidal field.  
Unfortunately, these models possess infinite mass per unit length as a result.

Observations suggest that
some molecular clouds may be wrapped by helical fields (Bally 1987; Heiles 1987).
There is also some observational evidence for helical fields in HI filaments towards 
the Galactic high latitude clouds (Gomez de Castro, Pudritz, \& Bastien 1997). 
In fact, helical fields represent the most general magnetic field
configuration allowed if cylindrical symmetry is assumed.
A few authors have previously modeled filamentary clouds with helical fields.  These models are similar to the
\Stod (1963) solution in that the magnetic pressure is proportional to the gas pressure, so that the density becomes a
re-scaling of the Ostriker (1964) solution.

Our analysis replaces the assumption of constant $\beta$ with the assumption of constant flux to mass loading for the 
poloidal (eg. Mouschovias 1976, Spitzer 1978, Tomisaka, Ikeuchi, and Nakamura 1988) and toroidal fields.  
We show that the magnetic field in this case has non-trivial effects on the density 
distribution, and if fact results in much better agreement with the available data.
We also explore the role of the EOS by constructing models using both an ``isothermal'' 
EOS, where the total (thermal plus non-thermal) velocity dispersion is assumed constant, and the pure logatrope
of McLaughlin and Pudritz (1996, hereafter MP96).  The effects of pressure truncation play an important role in our analysis.
By including a realistic range of external pressures, appropriate for the ISM, we show that the mass per unit length
of our models is significantly decreased from the untruncated value.
We also derive a new formulation of the virial equation
appropriate for the radial equilibrium of pressure truncated filamentary equilibria with helical fields.
We use this equation to compare our models with real filamentary clouds and to establish 
strong constraints on their allowed magnetic configurations.

How would helical fields arise?
All that is required
is to twist one end of a filament containing
a poloidal field, with
respect to the other end.  
Even if molecular filaments form with an initially axial magnetic field,
a helical field is plausibly generated by  
any kind of shear motion (such as subsequent oblique shocks, torsional \Alfven waves, etc.)
that twists the field lines.  

It is not the purpose of this paper to examine how helical fields could
be generated.  The main point of this work is that, 
having recognized that most molecular clouds are undoubtedly filamentary, magnetized, and truncated by 
an external pressure, it is of considerable importance to 
investigate equilibrium models of molecular  
clouds that contain quite general helical fields and pressure truncation.   
We employ two main approaches in our theoretical analysis.  Firstly, we derive a
general virial equation appropriate for pressure-truncated
filamentary molecular clouds, which we use to understand the roles of gravity, pressure, and
the magnetic field in the overall quasi-equilibrium of filamentary clouds. 
Secondly, we develop numerical MHD equilibrium models that can be compared with the
internal structure of real clouds.

Our virial analysis demonstrates that poloidal fields always help to support the gas against self-gravity,
while toroidal fields squeeze the gas by the
``hoop stress'' of their curved field lines.  Helical fields may either support or help
to confine the gas, depending on whether the poloidal or toroidal field component is dominant.
We show, in fact, that it is very difficult to understand observed clouds without
the notion of helical fields and the confining hoop stresses that they exert
upon their molecular gas.

Having found evidence for helical fields from our virial analysis, we construct numerical MHD models of filamentary
clouds in order to investigate the internal structure of models that are allowed by the data.
It is noteworthy that our isothermal 
models with helical magnetic fields always produce density profiles that fall off as $r^{-1.8}$ to $r^{-2}$, in excellent
agreement with the data.  
We show that the toroidal field component is responsible for the more realistic behaviour, and 
that purely poloidal fields result in density profiles that fall even
more rapidly than $r^{-4}$.
We also consider the pure 
logatrope of McLaughlin and Pudritz (1996) as a possible effective EOS for 
the gas.  We find that our logatropic models have somewhat more shallow density profiles, but 
many are also in good agreement with the existing data. 

A brief outline of our paper is as follows.  We first present the results of virial analysis of 
self-gravitating, pressure truncated, filamentary clouds
containing both poloidal and toroidal field (Section \ref{sec:virial}).  In Section \ref{sec:exact},
we follow this up with a detailed analysis of the equations of magnetohydrostatic equilibrium 
describing self-gravitating filaments and discuss
important analytic solutions to these.  A full numerical treatment
of the equations is given in Section \ref{sec:numerical} where we also
constrain our 3-parameter models with a wide variety of
filamentary cloud data.  We discuss these results in 
Section \ref{sec:Discussion} and summarize in Section \ref{sec:summary}.   

\section{Virial Analysis for Filamentary Molecular Clouds}
\label{sec:virial}

In Appendix A, we use the tensor virial theorem to construct a scalar form 
of the virial theorem appropriate for pressure truncated 
filamentary clouds containining arbitrary helical fields.
After carrying out the manipulations therein, we obtain
\be
0=2\int P d\V-2\Ps\V +\W + \M,
\label{eq:virial}
\ee
where the gravitational energy per unit length is given by
\be
\W=-\int \rho r\frac{\di\Phi}{\di r} d\V.
\label{eq:om1}
\ee
and $\M$ is the sum of all magnetic terms (including surface terms):
\be
\M=\frac{1}{4\pi}\int B_z^2 d\V - \left( \frac{\Bzs^2+\Bphis^2}{4\pi} \right) \V.
\label{eq:mdef}
\ee
This equation is appropriate for a  
non-rotating, self-gravitating, filamentary molecular cloud whose length greatly exceeds its radius.
For the remainder of this paper, all quantities written with a subscript $S$ 
are to be evaluated at the surface of the filament; thus we write that our filament is truncated 
by an external pressure $\Ps$ at radius $\Rs$. 
We further reserve calligraphic symbols for quantities evaluated per unit length; $\W$ is the gravitational 
energy per unit length
since there are no external gravitational fields and $\V$ is actually the volume per unit length, or 
cross-sectional area $\pi \Rs^2$, of the filament.  
As we shall now show, $\W$ can be evaluated exactly for a
filament of arbitrary internal structure and equation of state.
The mass per unit length $m$ of the filament is obstained 
by simply integrating the density over the cross-sectional area:
\be
m=2\pi\int r \rho(r) dr.
\label{eq:mass}
\ee
Poisson's equation in cylindrical coordinates takes the form
\be
\frac{1}{r}\frac{d}{dr}\left(r\frac{d\Phi}{dr}\right)=\fpg \rho.
\label{eq:poisson1}
\ee
By integrating, we find that the mass per unit length interior to radius $r$ can be written as 
\be
m(r)=\frac{1}{2G}r \left.\frac{d\Phi}{dr}\right|_r.
\ee  
Using this result in equation \ref{eq:om1}, the gravitational energy per unit length
can be transformed into an integral over the mass per unit length:
\be
\W=-2G\int_0^m m' dm'=-m^2 G.
\label{eq:W}
\ee
{\em It is remarkable that the gravitational energy per unit length takes on the same
value regardless of the equation of state, magnetic field, 
or internal structure of the cloud.}  The only requirements are those of
virial equilibrium and cylindrical geometry.  McCrea (1957) gave an approximate formula
for the gravitational energy per unit length as $\W=-a m^2 G$ 
(where $a$ is a constant of order unity) based on
dimensional considerations; thus, our exact result gives $a=1$ for all cylindrical
mass distributions.  

By considering a long filament of finite mass $M$ and length $L$, 
we find that the gravitational energy scales quite differently for
filaments and spheroids:
\bea
W_{cyl} &=& -\frac{GM^2}{L} \nn\\
W_{sphere} &=& -\frac{3}{5}a\frac{GM^2}{R}, 
\eea
where $a$ depends on the detailed shape and internal structure for spheroids.
It is of fundamental importance that the gravitational 
energy scales with radius for spheroids, but not for filaments.  
McCrea (1957) used this point to argue that filaments
possess stability properties quite contrary to those of spheroidal
equilibria.  For spheroids, which best describe molecular
cloud cores, the gravitational energy scales as $\sim R^{-1}$.
As long as the core is magnetically subcritical,
there always exists a critical external pressure beyond which
the gravitational energy must dominate over the pressure support.
The equilibrium is unstable to gravitational collapse
past this critical external pressure.
On the other hand, the gravitational energy of a filament is unaffected by a change in radius.
Thus, the gravitational energy remains constant during any radial contraction caused by increased external
pressure.  If the filament is initially in equilibrium, gravity can never be made to dominate by squeezing the filament; 
all hydrodynamic filaments initially in equilibrium are stable in the sense of Bonnor (1956) and Ebert (1955).

In Appendix B, we consider the Bonnor-Ebert stability of magnetized filaments.  Beginning with a 
discussion of uniform filaments, we show that a uniform filamentary cloud with a helical field, that is initially 
in a state of equilibrium, cannot be made to collapse radially by increasing the external pressure.
We also give a more general proof which extends the argument to non-uniform filaments of arbitrary
EOS.  Thus, we conclude that all filamentary clouds, that are initially in a state of equilibrium,
are stable against radial perturbations.

The virial theorem for filaments (equation \ref{eq:virial}) is best used 
to study the global properties of filamentary molecular clouds.  
It is useful to define the average density, pressure, and magnetic pressure within the cloud as
\bea
\rhoave &=& \frac{m}{\V} \nn\\
\Pave &=& \frac{\int_\V P d\V}{\V} \nn\\
\Pmagave &=& \frac{1}{8\pi\V}\int_\V \Bz^2 d\V.
\label{eq:averages}
\eea
Quite generally, we may write the effective pressure inside a molecular cloud as $P=\sigma^2\rho$, 
where $\sigma$ is the total velocity dispersion.
We emphasize that all of our models take $\sigma$ to represent the total velocity dispersion, including
both thermal and non-thermal components.  It is particularly important to note that when we describe an equation of state as 
``isothermal'', we really mean that the {\em total} velocity dispersion is constant.
The average squared velocity dispersion is defined simply as
\be
\langle \sigma^2 \rangle=\frac{\langle P \rangle}{\langle \rho \rangle}
=\frac{\int_\V \sigma^2 \rho d\V}{\int_\V \rho d\V},
\label{eq:sigsqave}
\ee
where the average has been weighted by the mass as in MP96.

With the above definitions, we easily derive a useful form of our virial equation (equation \ref{eq:virial}):
\be
\frac{\Ps}{\Pave}=1-\frac{m}{\mvir}\left(1-\frac{\M}{|\W|}\right),
\label{eq:MW}
\ee
where $\M$ and $\W$ are the total magnetic and kinetic energies per unit length defined in equations
\ref{eq:om1} and \ref{eq:mdef}, and $\mvir$ is the virial mass per unit length defined by
\be
\mvir=\frac{2 \sigsqave}{G}.
\label{eq:mvir}
\ee
We note that $\mvir$ is analogous to the the virial mass
\be
M_{vir}=\frac{5R\sigsqave}{G}
\ee
normally defined for spheroidal equilibria.
Using the definition of the 
average magnetic pressure given in equation \ref{eq:averages}, we may write the total magnetic energy as
\be
\M=2\left(\Pmagave-\Pmags\right)\V.
\label{eq:MCAL}
\ee
Using this result, along with the expression for the gravitational energy per unit length
(equation \ref{eq:W}) in equation \ref{eq:MW}, we obtain another useful for for our virial equation
after some algebraic manipulations:
\be
\frac{\Ps}{\Pave}=1-\frac{m}{\mvir}+ \left( \frac{ \Pmagave -\Pmags }{\Pave} \right),
\label{eq:virial2}
\ee
where $\Pmags$ is the total magnetic pressure evaluated at the surface of the cloud:
\be
\Pmags=\frac{\Bzs^2+\Bphis^2}{8\pi}.
\ee

Equation \ref{eq:virial2} makes two important points.  First of all, the poloidal
component of the magnetic field contributes to the magnetic pressure support of the cloud through
$\Pmagave$.
Secondly, the toroidal field enters into equation \ref{eq:virial2} only as a surface
term, through $\Pmags$, which helps to confine the cloud by  
the ``pinch effect'' well known in plasma physics.  

All magnetic fields, whether poloidal, toroidal, or of a more complex geometry, are associated with currents  
that flow within molecular clouds and the surrounding ISM.  For a filamentary cloud wrapped by a helical field,  
the toroidal field component implies the existence of a poloidal current that flows along the filament.
A natural question is whether a return current outside of the filamentary cloud completes the ``circuit'', or whether
the poloidal current connects to larger scale structures in the ISM.  The answer to this question will
likely depend on the mechanisms by which filaments form, which might be addressed by future analysis.
If the current returns as a thin current sheet flowing along the surface of the filament,
the toroidal field at the surface would be nullified, and so would its confining effects.  As we show in Section
\ref{sec:obs}, this would make the available data very difficult to understand, indeed.  However, if
the return current is diffuse and extended throughout the surrounding gas, as in the case of protostellar jets (Ouyed \& Pudritz 1997),
there would be a net magnetic confinement of the filament, which is consistent with the observations.

In Appendix C, we derive the virial relations for filamentary
molecular clouds analogous to the well known relations for spheroidal
clouds (Chi\`eze 1987; Elmegreen 1989; MP96).  We show that the two geometries
result in differences only in factors of order unity. 
Most importantly, we use our virial equation \ref{eq:virial} to show that Larson's laws (1981) 
are also expected 
for magnetized filamentary clouds of arbitrary EOS.  
The reader may consult Table \ref{tab:virialtable} to compare expressions for $m$, $R$, $\rhoave$, 
and $\Sigma$ for spheroidal and filamentary clouds.

\subsection{Unmagnetized Filaments}
\label{sec:unmagvir}
From equation \ref{eq:MW}, we see that unmagnetized clouds obey the following linear relation:
\be
\frac{\Ps}{\Pave}=1-\frac{m}{\mvir}.
\label{eq:unmag}
\ee
This equation is exact for any unmagnetized filamentary cloud in virial equilibrium regardless
of the underlying equation of state or details of the internal structure.  Since equation
\ref{eq:unmag} contains only quantities that are observable, we have
derived an important diagnostic tool for determining whether or not
filamentary clouds contain dynamically important ordered magnetic fields.

We can use equation \ref{eq:unmag} to obtain the critical mass per unit length $\mh$ for
unmagnetized filamentary clouds.  We consider a thought experiment in which mass is gradually 
added to a self-gravitating hydrostatic filament.  As the mass per unit length increases,
the compression due to self-gravity drives the filament to ever increasing internal pressures, while
the external pressure remains constant.  This process continues until the cloud is so highly compressed
that $\Ps/\Pave \rightarrow 0$, beyond which no physical solution to equation \ref{eq:unmag} exists.  
By equation \ref{eq:unmag}, this happens when $m=\mvir$; 
thus, the virial mass per unit length plays the role of the critical mass per unit length $m_h$ for
unmagnetized filamentary clouds.

For a prescribed EOS, this procedure leads to an unambiguous determination of the value of
$\mh$.  Depending on the EOS, the mass per unit length either
approaches $\mh$ asymptotically as $\Ps/\Pave \rightarrow 0$,
or achieves $\mh$ at some finite radius where $\Ps$ vanishes.

There is, however, one subtle point that needs to be made.  
For an isothermal equation of state, we can unambiguously write that $m_h=\mvir$.
However, the velocity dispersion varies with density for non-isothermal equations of state.
Thus, $\sigsqave$ and hence $\mvir$ (by equation \ref{eq:mvir}) may vary 
as the cloud is compressed by self-gravity.  The critical mass per unit length $m_h$
is the {\em final} value that $\mvir$ takes before radial collapse ensues, while
$\mvir$ is a quantity that applies equally well to non-critical states.

\subsection{Magnetized Filaments}
\label{sec:mag}
When there is a magnetic field present in a filamentary molecular cloud, the critical mass per unit length
$\mmag$ is significantly modified from the result obtained for unmagnetized clouds in the previous Section.
Using the same argument as presented above, a magnetized cloud achieves its critical configuration
when $\Ps/\Pave \rightarrow 0$:
\be 
\mmag=\frac{\mvir}{1-\M/\W}
\label{eq:mmag}
\ee
where $\M$ and $\W$ are the total magnetic and gravitational energies per unit length given by equations
\ref{eq:MCAL} and \ref{eq:W}.  We recall that $\M$ may be either positive or negative, depending on whether the 
poloidal or the toroidal field dominates the overall magnetic energy.  {\em In general, we find that poloidal
fields increase the critical mass per unit length beyond $\mvir$ for hydrostatic filaments,
while toroidal fields reduce the critical mass per unit length below $\mh$.}
Physically, the reason for this behaviour is that the poloidal field helps to support the cloud radially against 
self-gravity, thus allowing greater masses per unit length to be supported.  The opposite is true for 
the toroidal field component, which works with gravity in squeezing the cloud radially.

We may constrain the critical mass per unit length for a filamentary cloud if 
we have additional information regarding the strengths of the poloidal and toroidal field components.
For a nearly isothermal EOS, the magnetic critical mass per unit length is given by
\be
\mmag \approx \mh \left\{1+\left[ \frac{\Pmagave-\Pmags}{\Pave}\right]\right\}.
\ee
Since molecular clouds are in approximate equipartition between their magnetic and kinetic
energies (Myers and Goodman 1988a,b, Bertoldi and McKee 1992), $\Pmagave$ is not likely to greatly exceed $\Pave$.
Therefore, it is unlikely that the magnetic critical mass per unit length $\mmag$ would exceed the hydrostatic
critical mass per unit length $\mh$ by more than a factor of order unity.

\subsection{Surface Pressures on Molecular Filaments}
\label{sec:envelopes}
Molecular clouds are surrounded by the atomic gas of the interstellar medium (ISM).  
Like the molecular gas itself, the total pressure of the ISM is dominated by
non-thermal motions.
The external pressure is extremely important to our analysis since it 
both truncates molecular clouds at finite radius and helps to confine the clouds against their
own internal pressures (see equation \ref{eq:MW}).
Boulares and Cox (1990) have estimated the total pressure (with thermal plus turbulent contributions)
of the interstellar medium to be on the order of $10^4~K~cm^{-3}$.
However, some molecular clouds are associated with HI complexes, whose pressures are typically 
an order of magnitude higher than the general ISM (Chromey, Elmegreen, \& Elmegreen 1989).
Therefore, we will be absolutely conservative by assuming that the
external pressure on molecular clouds is in the range of $10^{4-5}~K~cm^{-3}$.  This
assumption almost certainly brackets the real pressure exerted on molecular clouds by imposing the total
(thermal plus turbulent) pressure of the ISM as a lower bound, and the pressure of large
HI complexes as the upper bound.  

While the above pressure estimate is appropriate for most filamentary clouds, which are truncated directly by the 
pressure of the surrounding atomic gas, we note that a second type of filament exists, in which a dense molecular 
filament is deeply embedded in a molecular cloud of irregular or spheroidal geometry.  The best example of this type
of filament is the $\int$-shaped filament in the Orion A cloud.  In such cases, the external pressure must be 
estimated using the density and velocity dispersion of the surrounding molecular gas.

\subsection{Comparison With Observations}
\label{sec:obs}
\begin{figure}
\hspace{\fill}
\psfig{file=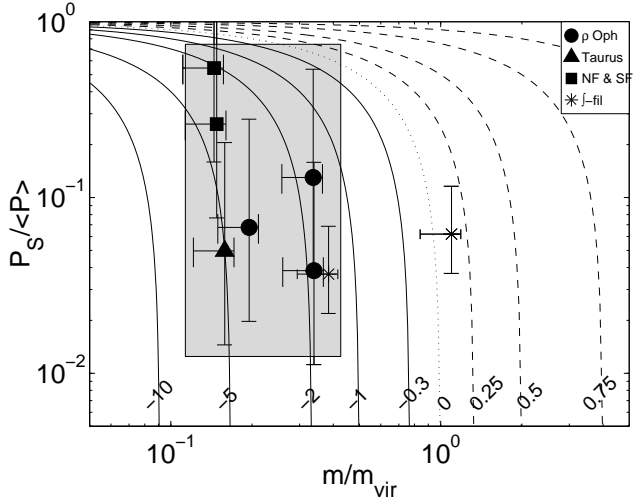,width=\linewidth}
\caption{Helical field models are compared with the observed properties of real filaments.
Curves are shown for various values of the virial parameter $\M/|\W|$.  Positive values, corresponding
to the dashed curves,
indicate that the poloidal field is dominant, while negative values, corresponding to solid curves,
indicate that the
toroidal field is dominant.  The dotted line represents all solutions that are neutrally affected
by the helical field (including the unmagnetized solution).  The $\int$-shaped filament appears twice, because we have used
two independent data sets in our analysis.}
\label{fig:observer}
\end{figure}

We have seen that the magnetic field affects the global properties of filaments only through the dimensionless 
virial parameter $\M/|\W|$.
The virial quantity $\M/|\W|$ provides a very
convenient index of whether a cloud is poloidally or toroidally dominated and to what degree.
For clouds with positive $\M/|\W|$, the net effect of the magnetic field is to provide support and
the field is poloidally dominated (cf. equation \ref{eq:MW}).  When $\M/|\W|$
is negative, the net effect of the field is confinement by the pinch of the
toroidal field, and the field is toroidally dominated.  Since $\M$ is directly
compared to the gravitational energy $|\W|$, the magnitude of our virial parameter provides an immediate
indication of the importance of the ordered field to the dynamics of the cloud.
In Figure \ref{fig:observer}, we have used equation \ref{eq:MW} to draw contours of constant $\M/|\W|$ as a function
of $m/\mvir$ and $\Ps/\Pave$.  The $\M/|\W|=0$ (dotted) line represents all helical field configurations,
including the unmagnetized special case, which
have a neutral effect on the global structure of the cloud.  Thus, we see that
the diagram is divided into poloidally dominated (dashed lines) and toroidally
dominated (solid lines) regions.

Since both $m/\mvir$ and $\Ps/\Pave$ are observable quantities, we can constrain our   
models by locating
individual filamentary clouds on this diagram.
However, we must first compute $m/\mvir$ and $\Ps/\Pave$ for each cloud by the following steps.
For each filament, we have found values for the mass, length, radius,
and average linewidth from molecular line observations in the literature
(see table \ref{tab:data} for references).
The mass per unit length $m$ is obtained by dividing the mass of the filament by
its length allowing for inclination effects by conservatively assuming all filaments to be oriented within
$45^\circ$ of the plane of the sky.
Since the emitting molecule (usually $\twelveCO$ or $\thirteenCO$) is always much
more massive than the the average molecule in molecular gas, the observed linewidth must
be corrected by applying Fuller and Myer's (1992) formula:
\be
\Delta v_{tot}^2=\Delta v_{obs}^2 +
8 \ln (2) kT\left(\frac{1}{\overline{m}}-\frac{1}{m_{obs}} \right),  
\label{eq:dv}
\ee
where $m_{obs}$ is the mass of the emitting species, $\overline{m}$  
is the mean mass of molecular gas, and $T$ is the kinetic temperature of
the gas.  For a normal helium abundance $Y=0.28$, $\overline{m}=2.33$.
We have assumed a temperature of $20~K$ for the gas; the exact
temperature chosen makes only a small difference in $\Delta v_{tot}$ since
the turbulent component of the linewidth always dominates on any scale larger
than a small core.  The velocity dispersion may be obtained from
equation \ref{eq:dv} by
\be
\sigma=\frac{\Delta v_{tot}}{\sqrt{8\ln 2}}.
\label{eq:sigma}
\ee
We identify $\sigma$ with $\sigsqave^{1/2}$ as defined
by equation \ref{eq:sigsqave}, since this velocity dispersion is obtained from an average linewidth for          
the entire cloud.  With $\sigsqave$ known, we compute $\mvir$ from
equation \ref{eq:mvir}, which directly gives us $m/\mvir$.
Obtaining the average radius $\Rs$ directly from maps, and hence the cross-sectional area $\cal V$,
the average density and internal pressure are then easily obtained using equations 
\ref{eq:averages} and \ref{eq:sigsqave}.  
All that remains to deduce $\M/|\W|$ from equation \ref{eq:MW}
is to estimate the external pressure.  Most of the filamentary clouds in our sample are surrounded 
by atomic gas.  Therefore, we conservatively assume the total external pressure to
be in the range $10^4-10^5~K~cm^{-3}$, as discussed in Section \ref{sec:envelopes}.
In fact, the only exception is the $\int$-shaped filament of Orion A, which is deeply embedded in molecular
gas.  In this case, we have estimated the external pressure from measurements of the density and linewidth in the
Orion A cloud (See table \ref{tab:reduceddata} for references).

\begin{table*}
\hspace{\fill}
\begin{minipage}{\linewidth}
\begin{tabular}{|l|l|c|c|c|c|l|l|} \hline
Cloud           &Region         &   M          &   L  &  $\Rs$  &  $\sigma$                 & Ref.    & Notes       \nn\\
                &               &$(\Msol)$     & (pc) &  (pc)   & $(km s^{-1})$             &         &             \nn\\
\hline

L1709           &Rho Oph        &  140         & 3.6  &  0.23   & 0.479                     & 2       & 1           \nn\\
L1755           &               &  171         & 6.3  &  0.152  & 0.526                     & 2       & 1           \nn\\
L1712-29        &               &  219         & 4.5  &  0.156  & 0.534                     & 2       & 1           \nn\\
DL 2 \footnote{Dark lane in Taurus including B18.  See Mizuno et al (1995) for a more detailed map.}
                &Taurus         & 600          & 6.4  & 0.5     & 1.08                      & 4       & 4           \nn\\
$\int$-fil. \footnote{The $\int$-shaped filament in Orion A.}
                &Orion          &  $5\x10^3$   & 13   &  0.25   & 1.41                      & 1       & 2,3         \nn\\
                &               &  $-$         & $-$  &  0.35   & 1.13                      & 5       &             \nn\\
NF \footnote{Northern filament in Orion (See reference).}
                &               & $1.55\x10^4$ & 87.3 & 2.25    & 1.54                      & 3       & 1           \nn\\
SF \footnote{Southern filament in Orion (See reference).}
                &               & $3.65\x10^4$ & 300  & 2.25    & 1.29                      & 3       & 1           \nn\\
\hline
\end{tabular}
\end{minipage}
\hspace{\fill}
\caption{We have compiled data on filamentary molecular clouds from several sources.
\protect\newline {\bf References:} 1. Bally (1987), 2. Loren (1989),
3. Maddalena (1986), 4. Murphy and Myers (1985), 5. Tatematsu et al. (1993)
\protect\newline
{\bf Notes}
1. Little star formation.  2. Dense cores, star formation.  3. Deeply embedded in Orion A cloud.  4. Associated stars.}
\label{tab:data}
\end{table*}

\begin{table*}
\hspace{\fill}
\begin{minipage}{\linewidth}
\begin{tabular}{|l|c|c|c|c|c|c|c|} \hline
Cloud           &m                      & $\mvir$               &  $m/\mvir$    &  $\Pave$              & $\Ps$                 & $\Ps/\Pave$   \nn\\
                & $(\Msol pc^{-1})$     & $(\Msol pc^{-1})$     &               &$(10^4K cm^{-3})$  &   $(10^4K cm^{-3})$       &               \nn\\
\hline
L1709           & 35.9                  & 107                   & 0.34          & 24.3                  & 3.2                   & 0.13          \nn\\
L1755           & 25.1                  & 129                   & 0.20          & 46.8                  & 3.2                   & 0.068         \nn\\
L1712-L29       & 45                    & 132                   & 0.34          & 82.4                  & 3.2                   & 0.038         \nn\\
DL 2            & 86.6                  & 547                   & 0.16          & 63.6                  & 3.2                   & 0.050         \nn\\
$\int$-fil.     & 355                   & 925                   & 0.38          & $1.77\x 10^3$         & 64.8
        \footnote{Determined from Bally's (1987) density estimate and the $\twelveCO$ linewidth given by Maddalena (1986).}     & 0.037         \nn\\
                & 647                   & 590                   & 1.1           & $1.05\x 10^3$         & 64.8                  & 0.062         \nn\\
NF              & 164                   & $1.11\x 10^3$         & 0.15          & 12.1                  & 3.2                   & 0.26          \nn\\
SF              & 112                   & 777                   & 0.15          & 5.8                   & 3.2                   & 0.55          \nn\\
\hline
\end{tabular}
\end{minipage}
\hspace{\fill}
\label{tab:reduceddata}
\caption{We have reduced the data of table \ref{tab:data} to obtain $m/\mvir$ and $\Ps/\Pave$ for each filament.
We assume an external pressure $\Ps$ of $10^{4.5\pm0.5}~K~cm^{-3}$ for all filaments except the $\int$-shaped filament of Orion A, which
is deeply embedded in molecular gas.  We also assume that all filaments are oriented within $45^\circ$ relative to the
plane of the sky.  We only give central values in the table, but the corresponding error bars are shown in figure \ref{fig:observer}.} 
\end{table*}

Figure \ref{fig:observer} demonstrates that most filamentary clouds reside in a part of parameter space where
\bea
0.11 &\appleq& m/\mvir \appleq 0.43 \nn\\
0.012 &\appleq& \Ps/\Pave \appleq 0.75, \nn\\
\label{eq:constraints}
\eea
which is indicated by the shaded box in Figure \ref{fig:observer}.
Thus, we find that filamentary clouds range considerably in their virial parameters.
{\em However, it is remarkable that most of the clouds in our small data set 
appear to reside in the part of the diagram where  $\M/|\W|<0$.
Thus, our virial analysis infers that the magnetic field in at least several 
filamentary clouds is probably helical and toroidally dominated.}
Gravity and surface pressure alone appear to be insufficient to radially bind the clouds in our sample.
While this means that filaments must be quite weakly bound by gravity,
we note that similar results have also been obtained by Loren (1989b) and BM92.

It is natural to wonder to what extent these conclusions could be affected by uncertainties in the observational
results.  The dominant sources of uncertainty in Figure \ref{fig:observer} are probably the uncertainties in mass per unit length
surface pressures.  However, we have assumed very conservative ranges for the surface pressures and inclination angles
of the clouds.  Therefore, we do not believe that observational uncertainties can account for the helical fields that are
required by our virial analysis. 
We also note that a more detailed model including rotation of the filament would necessarily lead to the same conclusion
of a helical field.  Since rotation would tend to support the cloud against gravity, even stronger toroidal fields would be
required to confine the gas.

\section{Exact MHD Models of Filamentary Structure}
\label{sec:exact}
The virial treatment of the previous section is perhaps the simplest
and most illuminating way to understand the physics and global properties 
of filamentary molecular clouds.  While the virial equations \ref{eq:virial} and 
\ref{eq:virial2} are convenient to use, and are in fact exact expressions of 
magnetohydrostatic equilibrium, the analysis can say nothing of the internal
structure of the clouds.  This is the advantage of the exact analytic 
and numerical models developed in this section.

\subsection{The Poloidal and Toroidal Flux to Mass Ratios}

\begin{figure}
\psfig{file=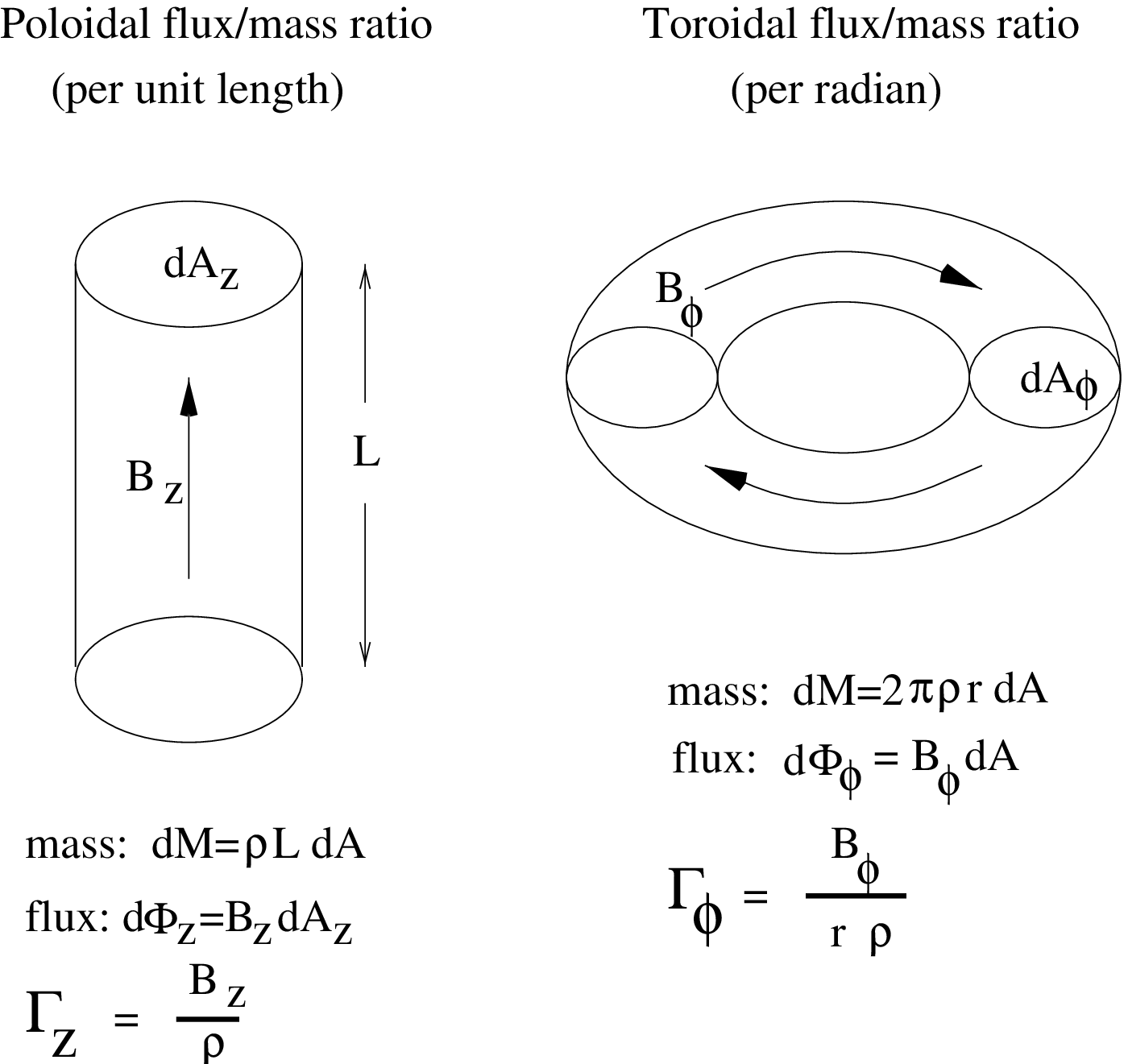,width=\linewidth}
\caption{A schematic illustration of the poloidal and toroidal flux to mass ratios introduced in
equations \ref{eq:Gz} and \ref{eq:Gphi}.}
\label{fig:Gamma}
\end{figure}

We postulate that the magnetic field structure corresponds to that of constant
poloidal and toroidal flux to mass ratios $\Gz$ and $\Gphi$. The meanings of the flux to mass ratios
are illustrated in Figure \ref{fig:Gamma}, and are
defined in the 
following way.  
Consider a bundle of poloidal field lines
passing through a small cross-sectional area of the filament $\delta\V$.  The
magnetic flux passing through the surface is $\Bz \delta\V$, while the mass per unit
length is $\rho\V$.  Thus, the ratio of the poloidal flux to the mass per unit length is
\be
\Gz=\frac{\Bz}{\rho}.
\label{eq:Gz}
\ee
Is there an analogous quantity for the toroidal component of the field?
In fact the toroidal flux has been defined and is commonly
used in plasma physics (Bateman 1978).  Here, we consider a
bundle of toroidal flux lines with cross-sectional area $\delta A$ that form
a closed ring of radius $r$ centred on the axis of the filament.  The mass enclosed by the ring
is $2\pi r \delta A$.  Thus, we may define the toroidal flux to mass ratio (per radian) as
\be
\Gphi=\frac{\Bphi}{r\rho}.
\label{eq:Gphi}
\ee
The simplest field configuration is that of constant
$\Gz$ and $\Gphi$.
We note that constant $\Gphi$ results naturally if a filament of constant $\Gz$ and length $L$
is twisted uniformly through an angle $\phi$.  Then
\be
\frac{\Bphi}{\Bz}=\frac{r\phi}{L},
\ee
which leads to the result
\be
\Gphi=\left(\frac{\phi}{L}\right)\Gz.
\ee
We shall always assume constant $\Gz$ and $\Gphi$ for the remainder of this paper.

\subsection{An Idealized Model: Uniform Magnetized Filaments}
\label{sec:uniform}

As an iside, it is illustrative to consider uniform filamentary clouds are affected by 
helical fields of constant $\Gz$ and $\Gphi$.  In this simple model, the 
magnetic field is uniform within the filament, but drops to zero outside.
By equation \ref{eq:Gphi}, the toroidal field $\Bphi$ increases as $\sim r$ within
the filament and falls off as $\sim r^{-1}$ in the external medium.  Thus, the 
toroidal field is associated with a constant poloidal current density within the filament.

With the assumption of constant density and the above definitions of $\Gz$ and $\Gphi$,
equation \ref{eq:virial2} can be expanded to give
\be
0=\left(\sigma^2-\frac{\Ps}{\rho}+\frac{\Gz^2\rho}{8\pi}\right)
-m\left(\frac{G}{2}+\frac{\Gphi^2}{8\pi^2}\right).
\label{eq:uniformcloud}
\ee
The critical mass per unit length is obtained by setting $\Ps=0$:
\be
\mmag=\frac{2\sigsq+\Gz^2\rho/4\pi}{G+\Gphi^2/4\pi^2}.
\ee

The effects of external pressure and the magnetic field are
transparent in this simple model.  Pressure and the poloidal field
cooperate in supporting the cloud.
On the other hand, the toroidal field
enters into equation \ref{eq:uniformcloud} in concert with gravity.  A filamentary cloud with
a helical field would be confined jointly by gravity, external pressure,
and the pinch of the toroidal field.
Without prior knowledge of the field
strength and direction (by molecular Zeeman and polarization observations), the cloud
may appear to be unbound by gravity alone.

\subsection{General Equations for Magnetized Filamentary Molecular Clouds}
\label{sec:formalism}

We consider the equilibrium structure of a 
non-rotating, self-gravitating molecular cloud with a helical field of 
constant flux to mass ratios $\Gz$ and $\Gphi$.  We consider two possible equations of
state for the gas: 1) the ``isothermal''
equation of state $P=\sigma^2\rho$ where $\sigma$ is the total
velocity dispersion and 2) the ``pure logatrope'' of MP96
given by $P/P_c=1+A\ln(\rho/\rho_c)$, where $P_c$ and $\rho_c$ are the central
(along the filament axis) pressures and densities, and $A$ is a constant.  
MP96 find $A\simeq 0.2$ for molecular cloud cores.  Although their analysis
was based only on cloud core data, we shall assume that the same value of $A$ 
might apply to filamentary clouds as well.  
We use these two equations of state because they probably bracket the true underlying equation
of state for molecular clouds;  MHD cloud turbulence probably results in an EOS
softer than isothermal (MP96; Gehman et al. 1996), while the pure logatrope is the softest EOS
to appear in the literature.

It is convenient to work in dimensionless units where density and 
pressure are
scaled by their central values $\rho_c$ and $P_c$.
We further define the central velocity dispersion by
\be
\sigma_c^2=\frac{P_c}{\rho_c}.
\ee
A natural radial scale is then given by
\be
r_0^2=\frac{\sigma_c^2}{4\pi G \rho_c},
\label{eq:r0scale}
\ee
which defines the effective core radius of the filament.
Finally, we may define natural scales for the mass per unit length 
and magnetic field:
\bea
m_0 &=& r_0^2 \rho_c=\frac{\sigma_c^2}{4\pi G} \nn\\
B_0 &=& P_c^{1/2}.
\label{eq:scales}
\eea
Thus, all quantities are written in dimensionless form as follows:
\bea
{\tilde r} &=& r/r_0 \nn\\
{\tilde \rho} &=& \rho/\rho_c \nn\\
{\tilde m}  &=& m/m_0 \nn\\
{\tilde P} &=& P/P_c \nn\\
{\tilde \sigma} &=& \sigma/\sigma_c \nn\\
{\tilde \Phi} &=& \Phi/\sigma_c^2 \nn\\
{\tilde \Bz} &=& \Bz/B_0 \nn\\
{\tilde \Bphi} &=& \Bphi/B_0
\eea
Hereafter, we will only ever refer to $\Gz$ and $\Gphi$
in their dimensionless forms:
\bea
{\tilde \Gz} &=& \sqrt{\frac{\rho_c}{\sigma_c^2}} \left(\frac{\Bz}{\rho}\right) \nn\\
{\tilde \Gphi} &=& \frac{1}{\sqrt{\fpg}} \left(\frac{\Bphi}{r\rho}\right)
\label{eq:scalings}
\eea
\noi For brevity, we will drop the tildes for the remainder of this section
and the next (except for where ambiguity would result); all
quantities hereafter are understood to be written in dimensionless form unless
otherwise stated.
 
Our basic dimensionless 
equations are those of Poisson
\be
\frac{1}{r}\dr\left(r\dr\Phi\right)=\rho
\label{eq:poisson}
\ee
and magnetohydrostatic equilibrium 
\be
\dr\left(P+\frac{\Bz^2}{8\pi}\right)+\rho\dr\Phi+\frac{1}{r^2}\dr\left(\frac{r^2\Bphi^2}{8\pi}\right).
\label{eq:equilibrium}
\ee
In Appendix D, we construct the mathematical framework to solve these equations numerically for both
isothermal and logatropic equations of state.  We show that a solution to the dimensionless equations is 
characterized by three parameters, namely the flux to mass ratios $\Gz$ and $\Gphi$ defined by equations
\ref{eq:Gz} and \ref{eq:Gphi}, and a third to specify the (dimensionless) radius of pressure truncation.
We express this third parameter as a concentration parameter defined by $C$ defined as  
\be
C=\log_{10}\left(\frac{\Rs}{r_0}\right),
\label{eq:Cdef}
\ee
where $r_0$ is the radial scale defined by equation \ref{eq:scalings}.
We note that our definition of
$C$ is analogous to the concentration parameter $C=\log_{10}(r_t/r_0)$ defined for King models
of globular clusters (See Binney \& Tremaine 1987).
Our concentration parameter differs only in that the tidal radius $r_t$ is replaced by the pressure truncation radius,
and our $r_0$ is smaller by a factor of 3.
While we use $C$ primarily as a theoretical parameter, we note that it is in principle observable.

\subsection{Analytic Solutions}
\label{sec:analytic}

Before discussing numerical solutions, we derive
a few special solutions that can be expressed in closed analytic form.  
Specifically, we discuss the unmagnetized isothermal 
solution that was found by Ostriker (1964) (a brief derivation is given in Appendix D.).
We note that that this solution is a special case of a more general magnetized solution
obtained by \Stod (1963); for brevity, we shall refer to this solution as the Ostriker solution for
the remainder of this paper.
We also find a singular
solution for logatropic filaments.  It is unlikely that either of these 
special solutions describe real filaments, which are probably magnetized
and non-singular, but they do serve as important benchmark results to
compare with our more elaborate magnetized models.

\subsubsection{The Ostriker Solution: Unmagnetized Isothermal Filaments}
\label{sec:unmagiso}
The analytic solution for the special case of an unmagnetized
isothermal filament is easily obtained using the mathematical framework
in Appendix D.  It was first given by Ostriker (1964):
\be
\rho=\frac{\rho_c}{(1+r^2/8 r_0^2)^2},
\label{eq:rhoost}
\ee
where we have restored the dimensional units.  We note that the density decreases
as $\sim r^{-4}$ at large radii.  That such steep density
profiles have not been observed could be explained by three possibilities:
1) molecular clouds are not isothermal.  A softer EOS would give a less steeply
falling density at large radius;
2) real clouds contain dynamically important magnetic fields that modify the
structure of the filament at large radius;
3) real filaments are always truncated by external pressure.  If the filament
is truncated before the $\sim r^{-4}$ envelope is reached, such steep behaviour would not be
observed.
We demonstrate in Section \ref{sec:helix} that either of possibilities 1) or 2)
can explain the observed properties of molecular clouds.

\subsubsection{Singular Logatropic Filaments}
\label{sec:singularlog}
Although we have been unable to find the analogue of the Ostriker solution for
the logatropic EOS, we have been successful in finding a singular solution.
For this model, we reinterpret $\rho_c$ as the density
at some fiducial radius.  
We postulate a power law solution of the form 
\be
\rho \propto r^\alpha,
\ee
and find that a solution can only be obtained if $\alpha=-1$.  The final solution with
dimensional units restored is
\be
\frac{\rho}{\rho_c}=\sqrt{A} \left(\frac{r}{r_0}\right)^{-1}.
\label{eq:SLC}
\ee
It is useful to compare our solution with the singular logatropic sphere found by MP96: 
\be
\frac{\rho_{sphere}}{\rho_c}=\sqrt{2 A} \left(\frac{r}{r_0}\right)^{-1},
\label{eq:SLS}
\ee
where we have rewritten their solution using our definition of $r_0$.
(Their definition of $r_0$ differs from ours
by a factor of 3.  Our definition is the customary choice for filaments.).
It is remarkable that both singular logatropic spheres and filaments obey precisely
the same power law.

\section{Numerical Solutions}
\label{sec:numerical}
We now turn our attention to numerical solutions of equation \ref{eq:numsys}
using various values of the flux to mass ratios $\Gz$ and $\Gphi$.
Many of the solutions are shown out to very
large radius but may be truncated to reproduce any desired value of the concentration
parameter $C$ defined in equation \ref{eq:Cdef}.

\subsection{Numerical Results}
\label{sec:numres}
We have shown in Section \ref{sec:obs} that many filamentary clouds are probably wrapped by helical magnetic fields.
However, before considering the most general case of helical fields in Section \ref{sec:helix}, 
we first separately consider the effects of poloidal and toroidal magnetic fields in Section \ref{sec:pure}.
We note that purely poloidal fields are not allowed by our virial analysis, and purely toroidal fields are probably unrealistic.
Neverthess, this is the best way to understand the roles of each field component in our more general helical field models.

Equation \ref{eq:numsys} gives the set of differential equations that we integrate to 
produce our models.  The integration was done in a straightforward manner, using a standard Runge-Kutta method.

\subsubsection{Models With Purely Poloidal and Toroidal Fields}
\label{sec:pure}

\begin{figure*}

\begin{minipage}{0.49\linewidth}
\psfig{file=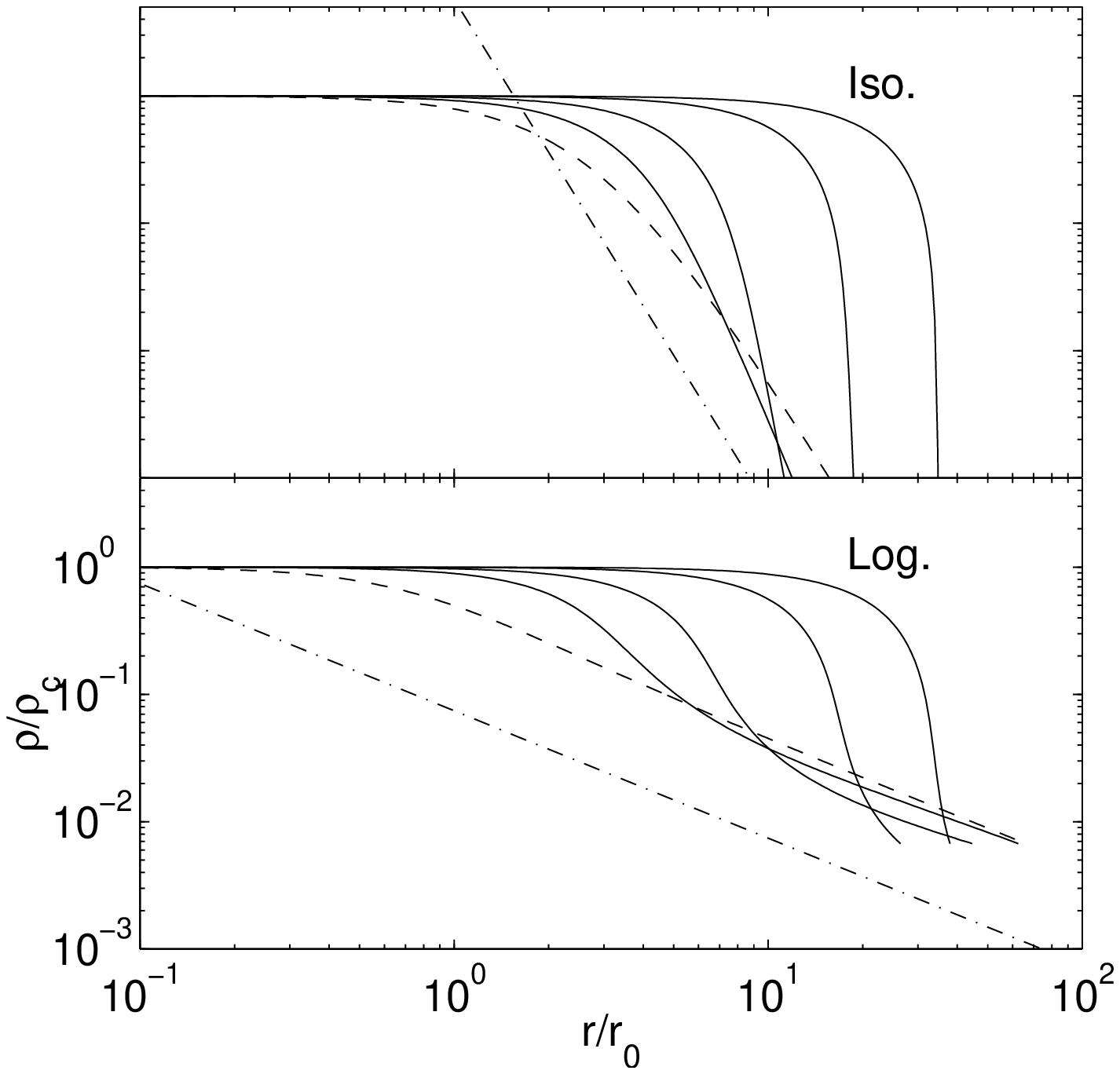,height=.31\textheight}
\end{minipage}
\begin{minipage}{0.49\linewidth}
\psfig{file=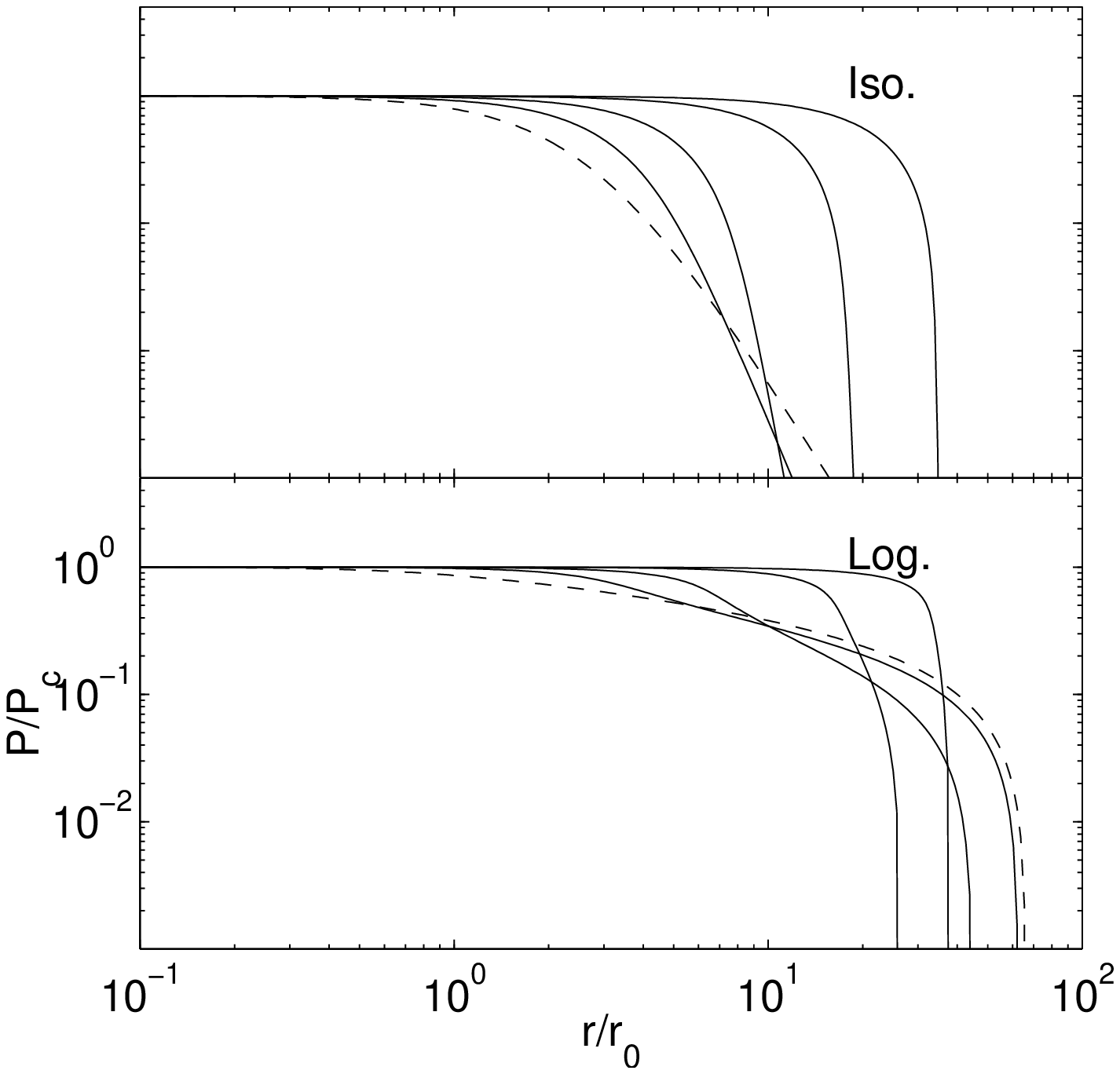,height=.31\textheight}
\end{minipage}
   
\begin{minipage}{0.49\linewidth}
\psfig{file=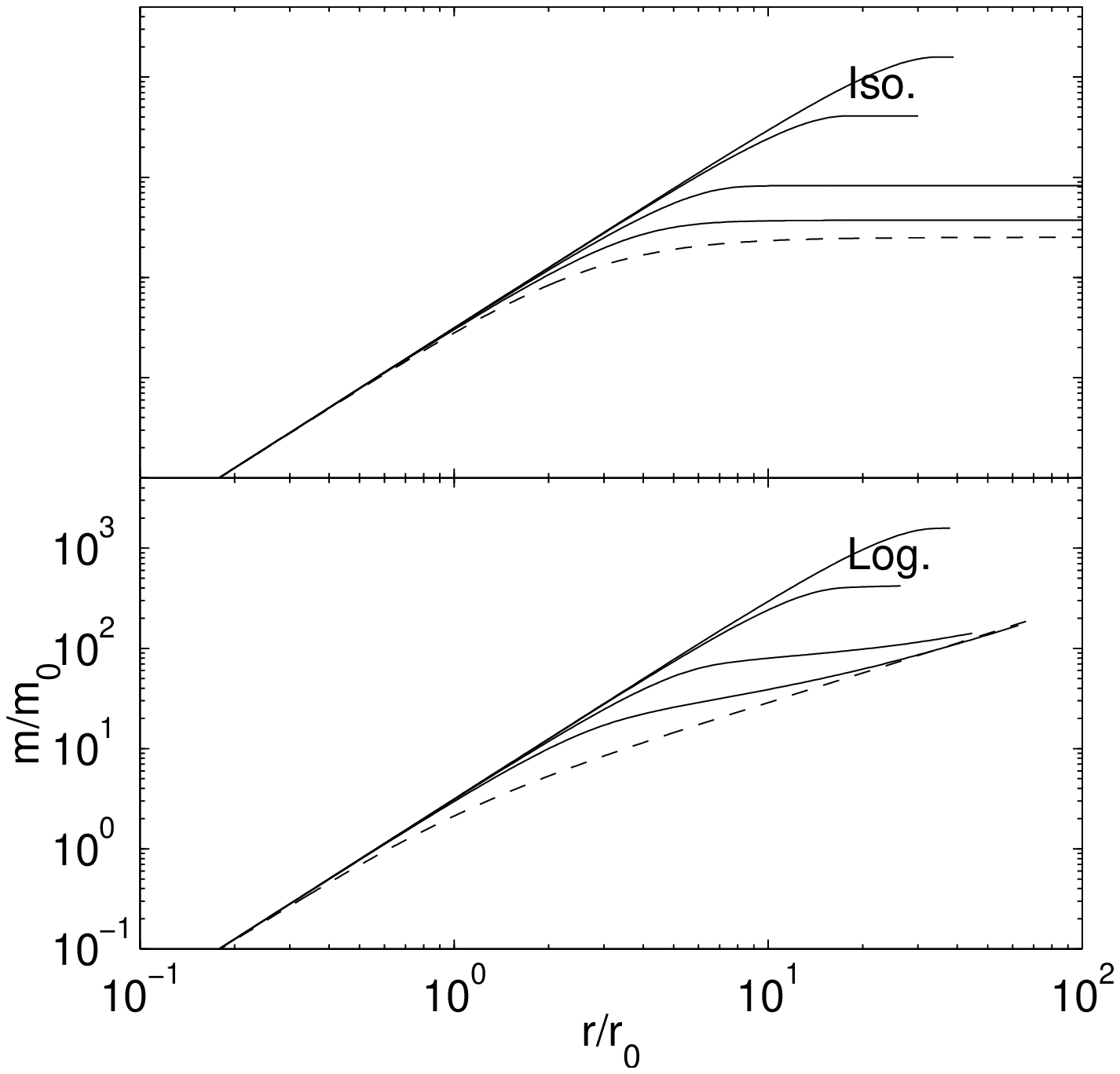,height=.31\textheight}
\end{minipage}
\begin{minipage}{0.49\linewidth}
\psfig{file=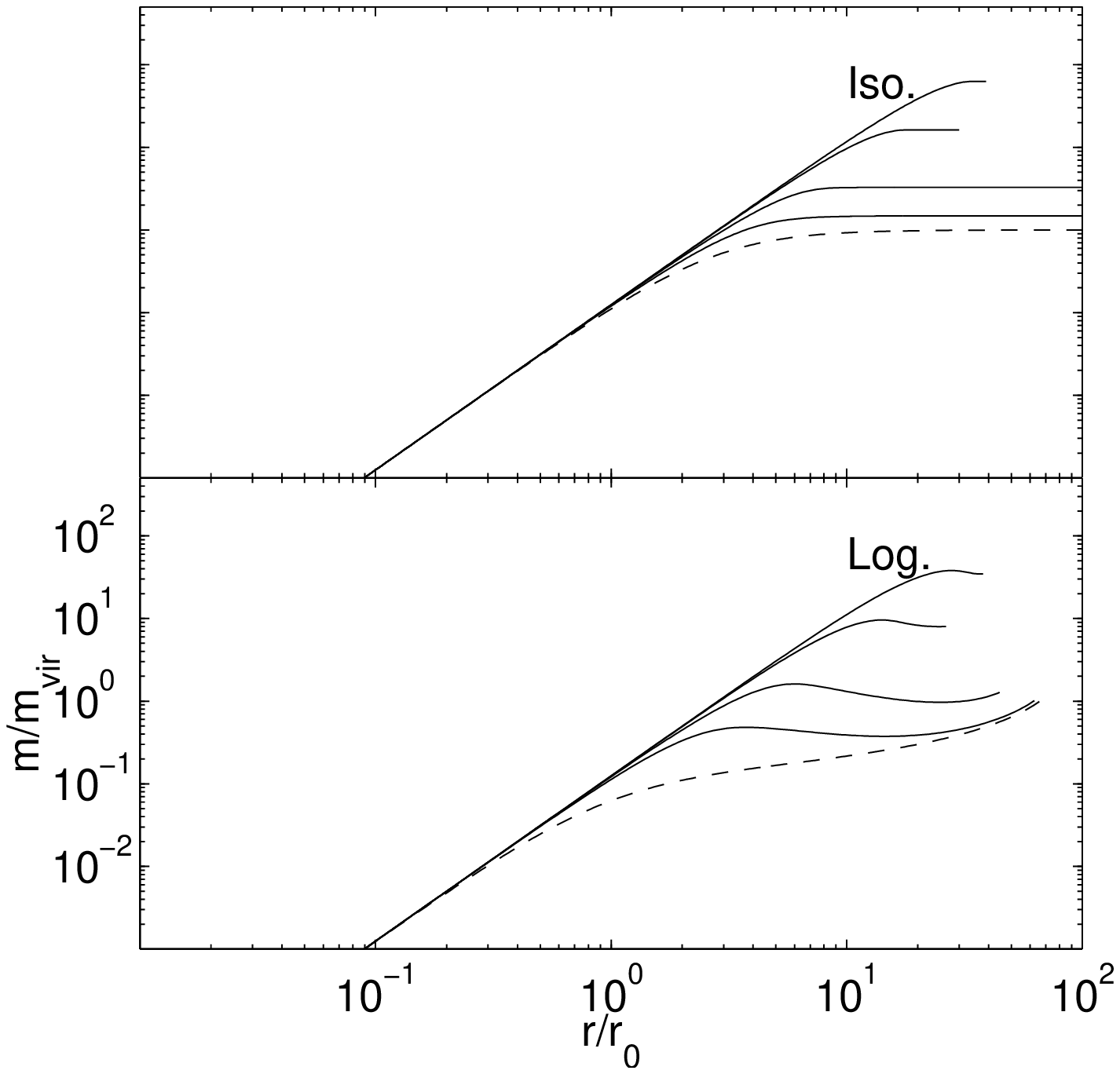,height=.31\textheight}
\end{minipage}

\begin{minipage}{0.49\linewidth}
\psfig{file=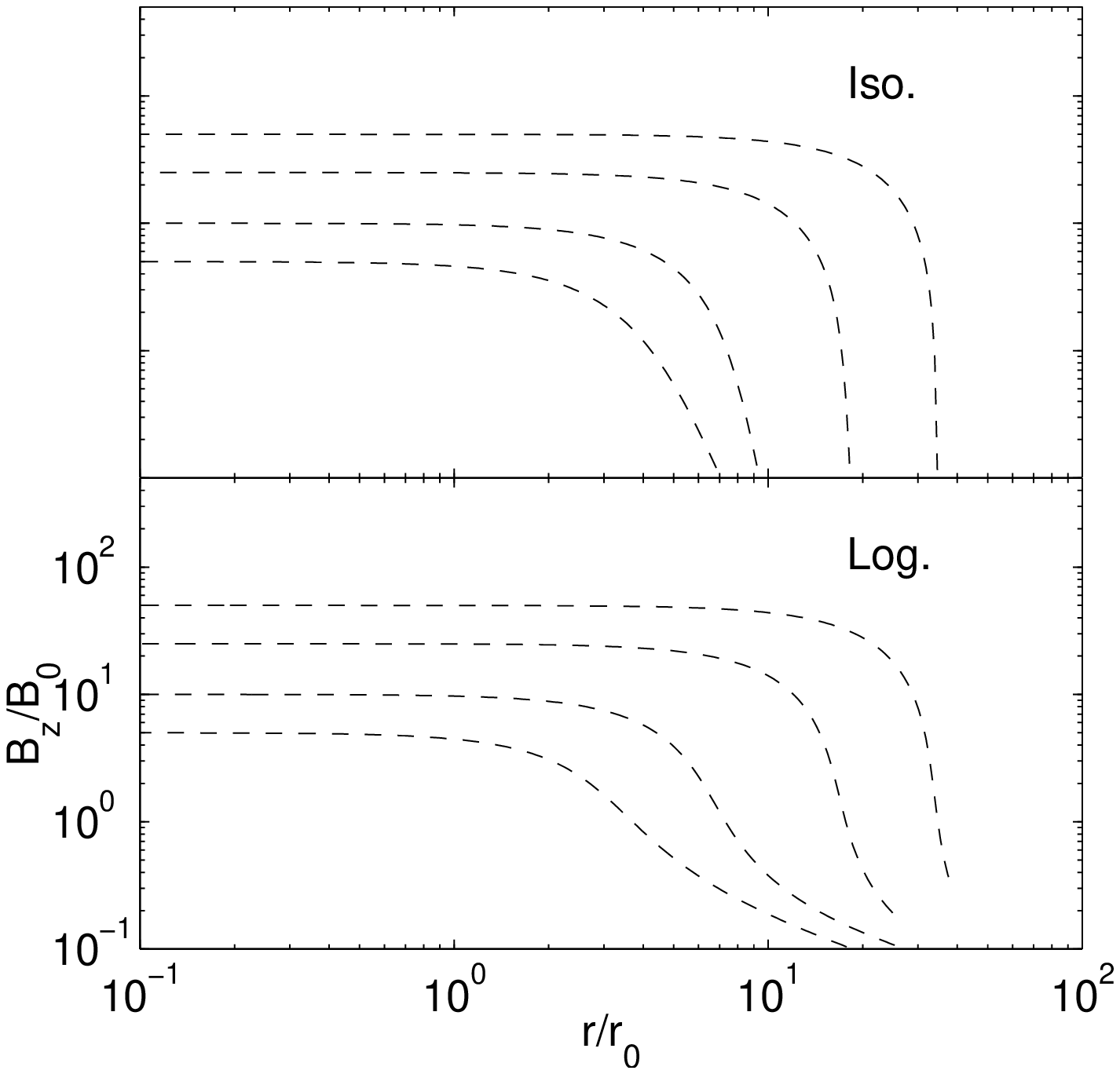,height=.31\textheight}
\end{minipage}
\begin{minipage}{0.49\linewidth}
\psfig{file=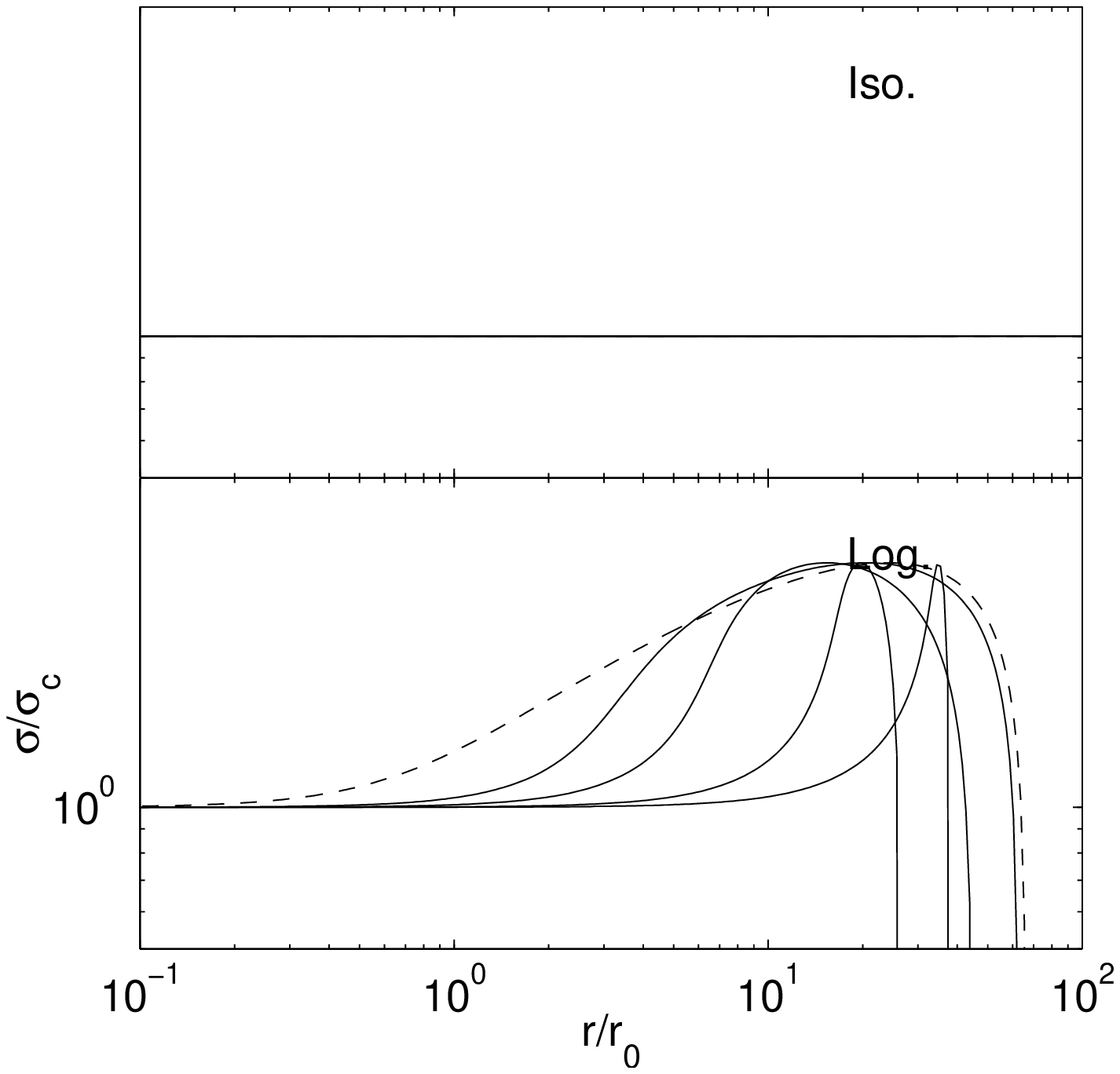,height=.31\textheight}
\end{minipage}

\caption[]{Isothermal and logatropic filaments with purely poloidal magnetic field:
$\Gz=0$ (dashed line), $5, 10, 25$ and $50$.  The dot-dashed lines
represent the $r^{-4}$ density stucture of the Ostriker solution
at large radius and the $r^{-1}$ behaviour of the singular logatropic solution.}

\label{fig:structure1}
\end{figure*}

\begin{figure*}

\begin{minipage}{0.49\linewidth}
\psfig{file=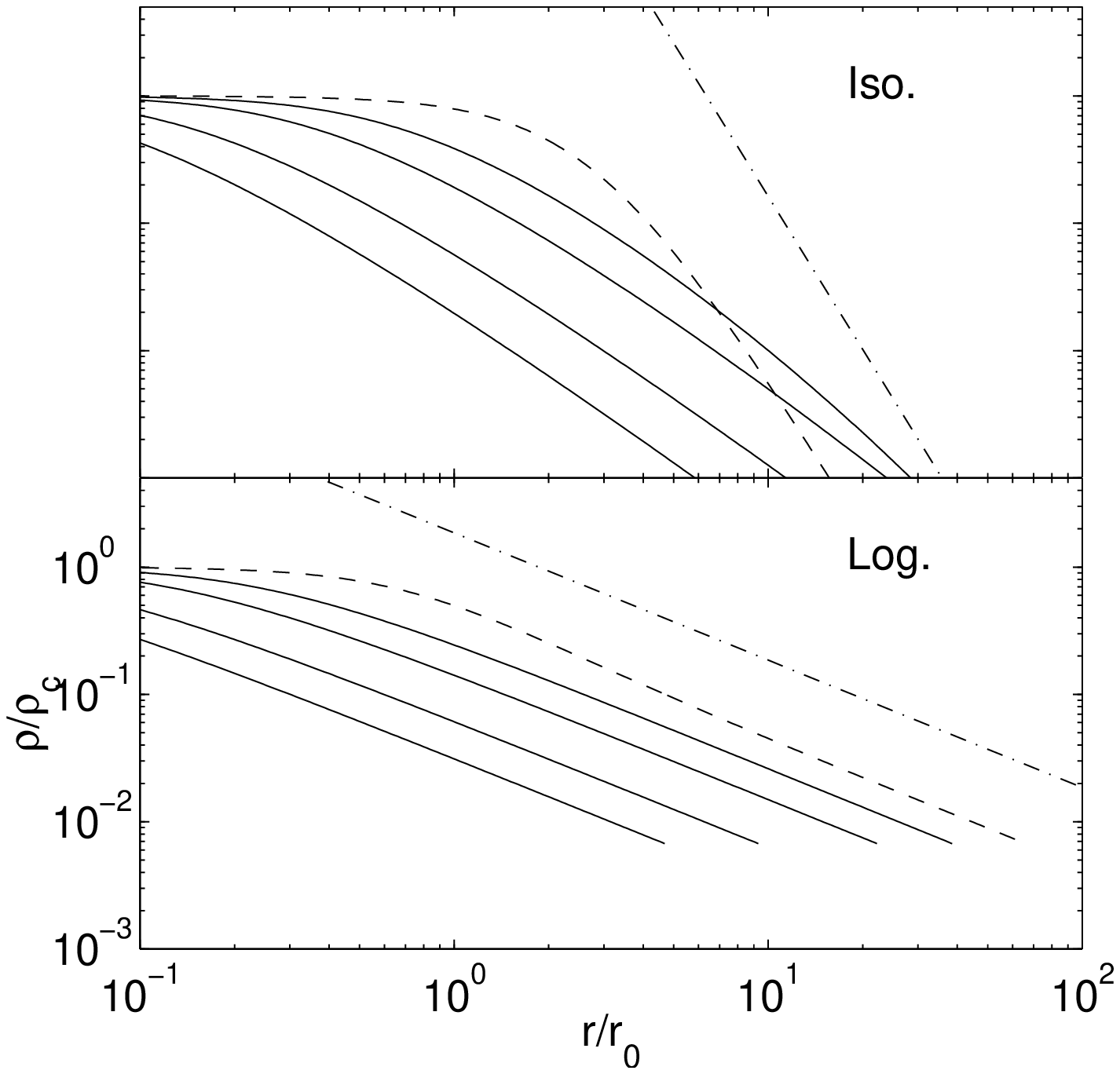,height=.31\textheight}
\end{minipage}
\begin{minipage}{0.49\linewidth}
\psfig{file=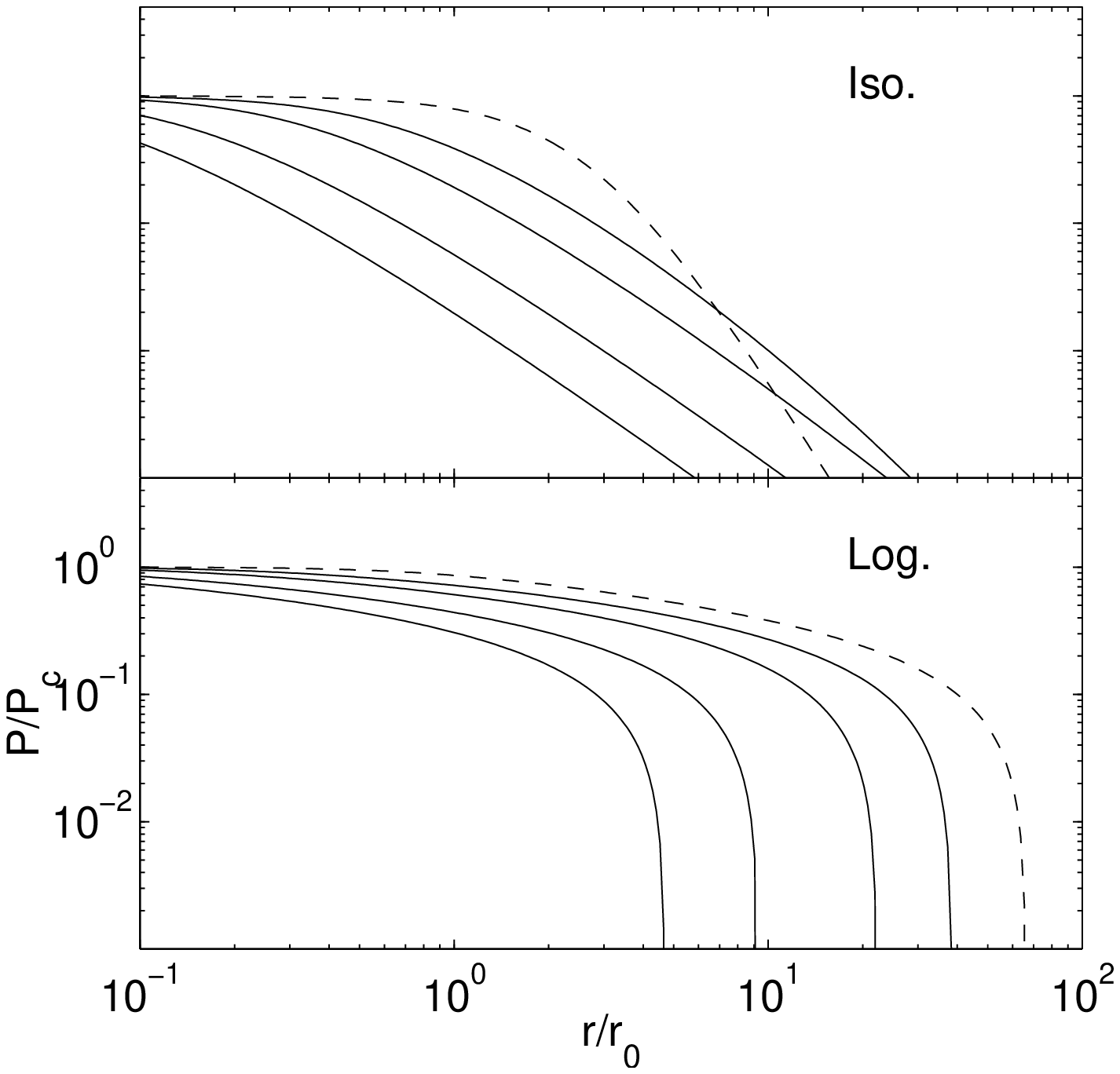,height=.31\textheight}
\end{minipage} 

\begin{minipage}{0.49\linewidth}
\psfig{file=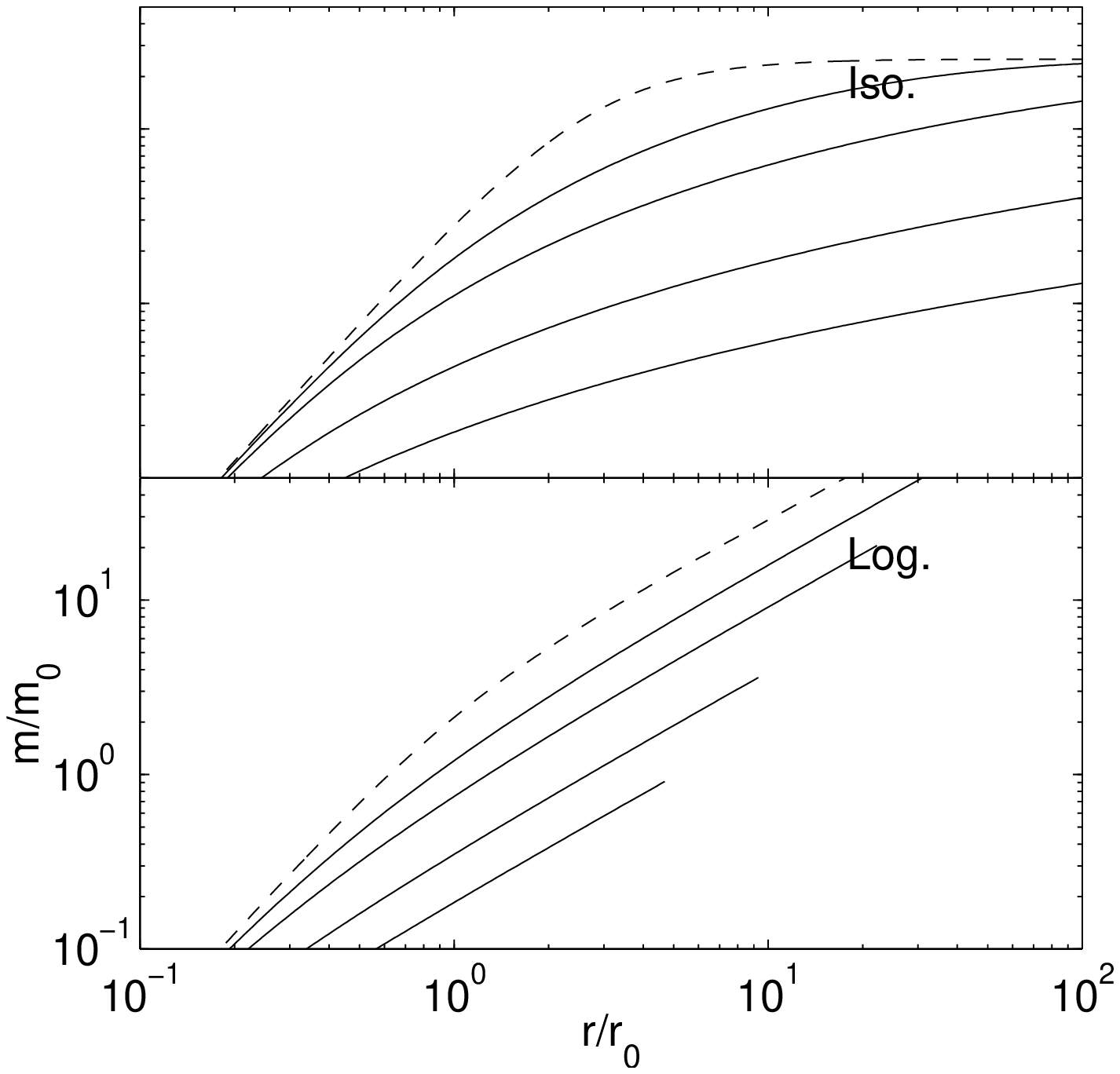,height=.31\textheight}
\end{minipage}
\begin{minipage}{0.49\linewidth}
\psfig{file=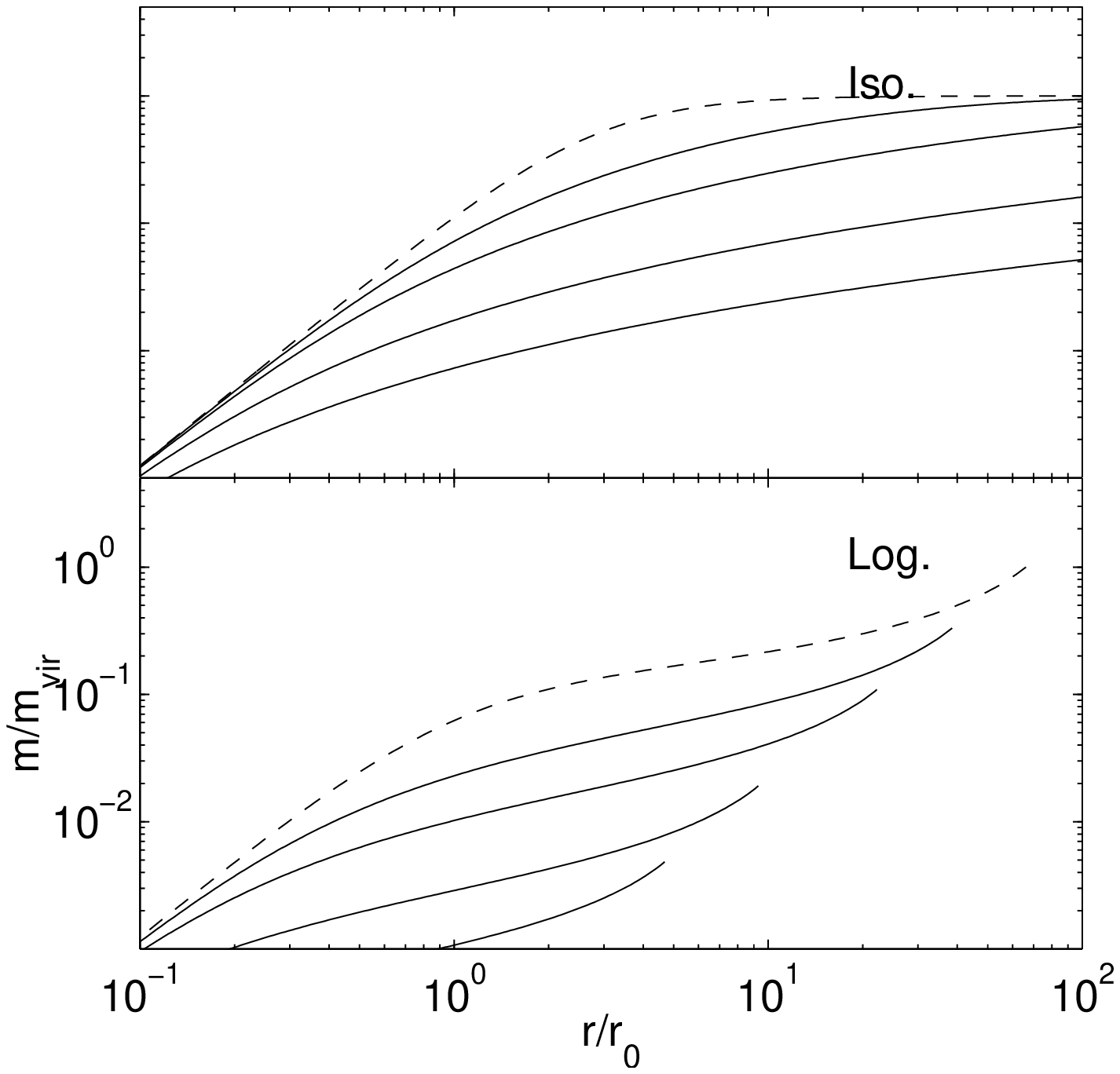,height=.31\textheight}
\end{minipage}

\begin{minipage}{0.49\linewidth}
\psfig{file=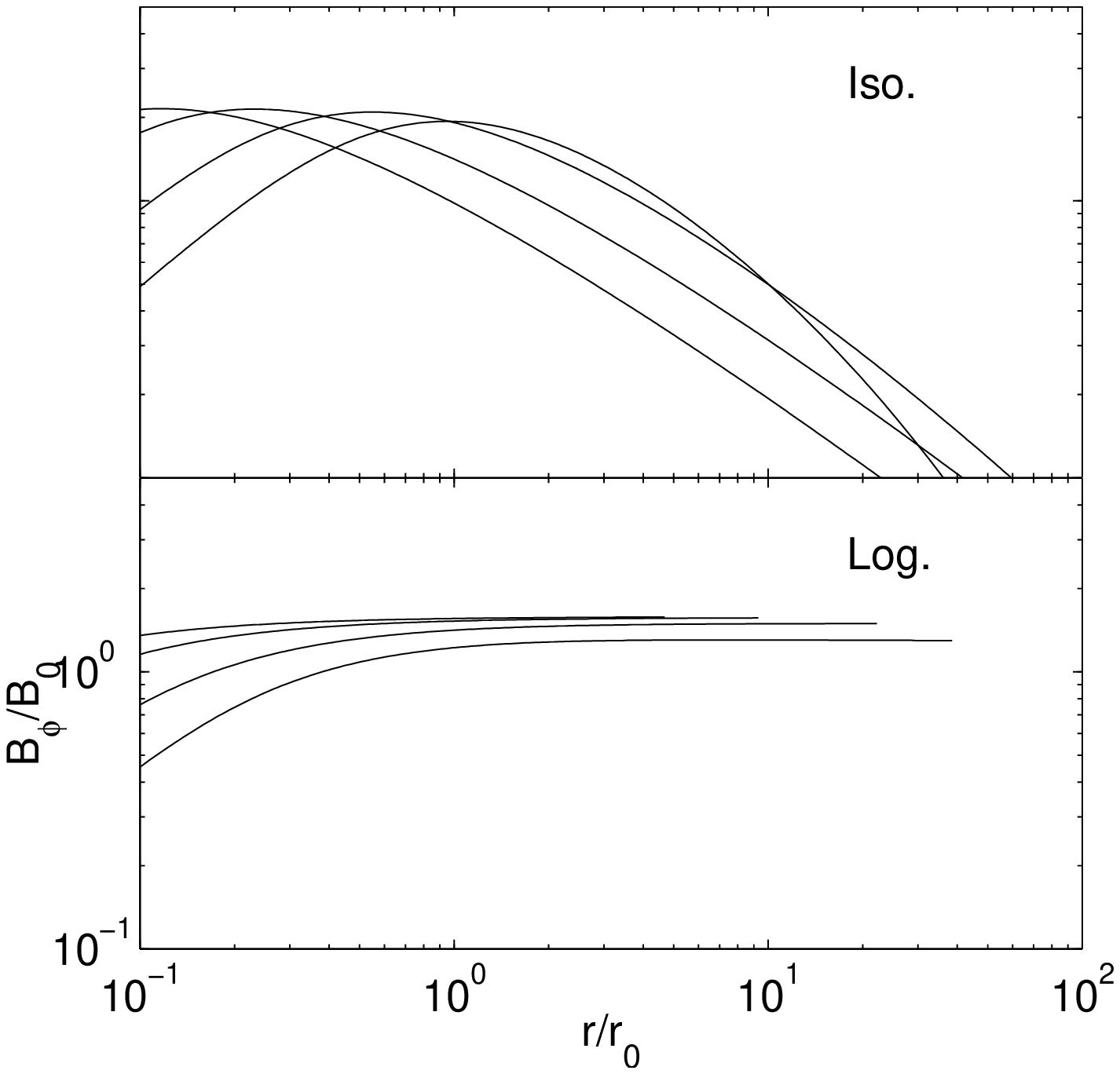,height=.31\textheight}
\end{minipage}  
\begin{minipage}{0.49\linewidth}
\psfig{file=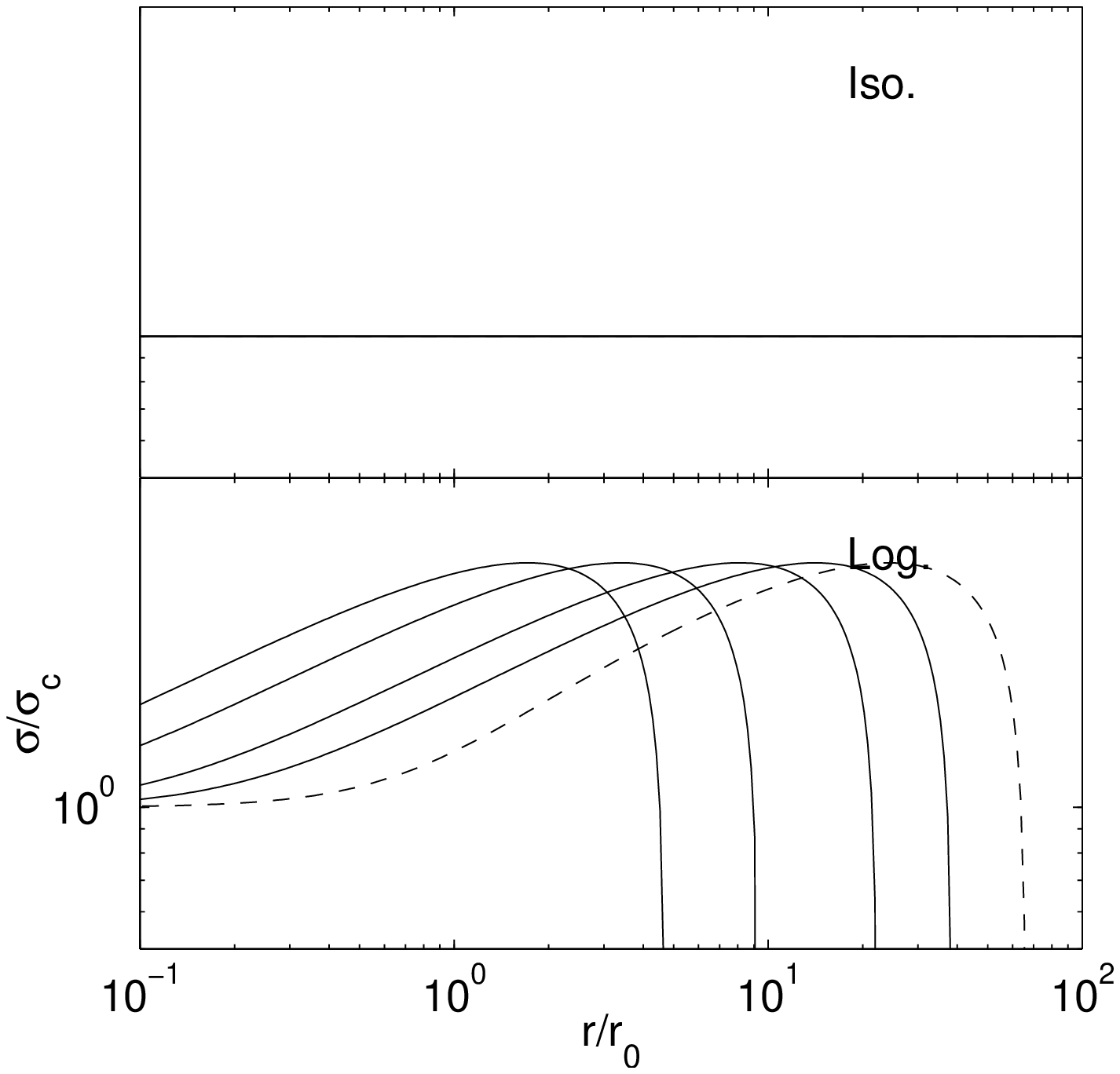,height=.31\textheight}
\end{minipage}

\caption[]{Isothermal and logatropic filaments with purely toroidal magnetic field:
$\Gphi=0$ (dashed line), $5, 10, 25$ and $50$.  The dot-dashed lines have the same 
meaning as in Figure \ref{fig:structure1}.}

\label{fig:structure2}
\end{figure*}

Figures \ref{fig:structure1} and \ref{fig:structure2}
show the density and pressure profiles, the magnetic structure, and the 
mass per unit length for isothermal
and logatropic filaments threaded by a purely poloidal field.  We have also included the velocity dispersion
and average velocity dispersion (given by equation \ref{eq:sigsqave}) for the
logatropic equation of state.

On each set of figures we have shown the density, pressure, and velocity dispersion
structure of the unmagnetized solutions with dashed lines.  For isothermal solutions,
we have also drawn
a line representing the asymptotic $r^{-4}$ behaviour of the Ostriker solution.
Similarly, the $r^{-1}$ singular solution has been included on density profiles
for logatropic filaments.
These power laws are meant as a guide
in interpreting the asymptotic behavior of the solutions;
we find that they are obeyed at large
radius for all unmagnetized filaments.

Comparing the unmagnetized solutions (dashed lines) in Figures \ref{fig:structure1} and \ref{fig:structure2},
we observe that the density
profile of the unmagnetized logatropic filament is slightly more centrally concentrated
than the isothermal Ostriker solution, but falls off much less steeply at large
radius.  These figures also show that
isothermal and logatropic filaments both tend to finite mass per unit length, although they differ in that  
isothermal filaments approach the critical mass per unit   
length only asymptotically as their radii tend to infinity.
As discussed in Section \ref{sec:virial}, this limit represents the critical mass per unit length $m_h$ (see 
Section \ref{sec:unmagvir}) beyond which no equilibrium is possible.  For the isothermal filament, it is easy to
show analytically (from equation \ref{eq:rhoost}) and we verify numerically that $m_h=8\pi m_0$, where $m_0$
is the mass scale defined by equation \ref{eq:scales}.  For the unmagnetized logatropic filament, we
find numerically that $m_h=185.8 m_0$.
Logatopic filaments can support a greater mass per unit length for equivalent central velocity
dispersion $\sigma_c$.  This is easily understood since 
the average velocity dispersion $\sigsqave^{1/2}$ always exceeds the central value offering more
turbulent support to the filament.

Perhaps the most notable feature of logatropic filaments is that they
``self-truncate'' at finite radius and density $\rho=\rho_c\exp{(-1/A)}$ where
the velocity dispersion and pressure vanish.
The logatropic EOS is designed to have a nearly isothermal
core and a rising velocity dispersion outside of the core radius.  At some point,
however, the velocity dispersion turns over and falls to zero.
The velocity dispersion could of course never vanish in a real cloud since
all real clouds are truncated by finite external pressure.
Whether the region of outwardly falling velocity dispersion
actually falls within the pressure truncation radius in fits to real clouds
is addressed in Section \ref{sec:best}, where we attempt to constrain our models 
using the observational results of table \ref{tab:reduceddata}.

Because of the way in which we have defined $\Gz$ (equation \ref{eq:Gz}), $\Bz$ is exactly
proportional to the density.  The toroidal field, however, shows a more
interesting structure; $\Bphi$ always vanishes along the axis of the
filament, as it must for the field to be continuous across the axis.  
We note that the logarithmic radial scale of Figure \ref{fig:structure2} makes the vanishing of $\Bphi$ 
at the axis difficult to see in some cases.
It also is found to
decay at large radius for isothermal filaments since
$\Bphi=\Gphi r\rho \propto r^{-3}$.  Hence, there is a single maximum in the toroidal
field structure.  
For logatropic filaments, the $r^{-1}$ asymptotic behaviour of the density implies that
$\Bphi=\Gphi r \rho$ tends to a constant value at large radius.  Thus, the toroidal field in logatropic filaments
lacks the local maximum found for isothermal filaments.

The effects of poloidal and toroidal magnetic fields on the density structure are apparent in Figures \ref{fig:structure1}
and \ref{fig:structure2}.  For either EOS, the poloidal magnetic field supports the cloud and causes it to 
be more extended radially than the corresponding unmagnetized filament.  Toroidal magnetic fields, on the other hand, 
pinch the filament to smaller radial extent.   Thus, we find our numerical results to be in agreement with the results of our virial 
analysis in Section \ref{sec:virial}.  
 
{\em It is significant that purely poloidal fields always steepen the outer density
profile, while toroidal fields make it more shallow.}
For isothermal filaments, poloidal fields always result in density profiles that are steeper
than the $r^{-4}$ behaviour of the Ostriker solution; this is true of even logatropic filaments 
when the field is of sufficient strength.  These steep density profiles
have never been observed, so purely poloidal fields do not match
the data.    
Thus, it seems likely that the field must have a toroidal
component if a realistic density profile is to be achieved.

We find that the magnetic field has a dramatic effect on the critical mass per unit length $m_{mag}$ of the cloud.
For either EOS, a poloidal magnetic field increases $m_{mag}$, since the poloidal field acts to support the cloud
against self-gravity.  The toroidal magnetic field works with gravity, thus decreasing the maximum mass per unit
length that can be supported.  These conclusions are in agreement with our virial results from Section \ref{sec:virial}.

From Figures \ref{fig:structure1} and \ref{fig:structure2}, we 
note that all isothermal filaments that are unmagnetized or 
contain a purely toroidal field tend to the same mass per unit length
$m_h$.  This is easily explained by our virial equation \ref{eq:virial2} 
since the toroidal field always tends to zero at large radius for isothermal
filaments.  The critical mass per unit length
$m_{mag}$ clearly cannot be affected, since the toroidal field only enters the virial equation through
its surface value.  This is not the case for logatropic filaments
because $\Bphi$ tends to a constant value at large radius.

\subsubsection{Helical Field Models}
\label{sec:helix}

In Section \ref{sec:obs}, we provided evidence based on our virial analysis that filamentary clouds
likely contain toroidally dominated helical magnetic fields.  At this point, we shall take a
further step by comparing our exact MHD models with the observed properties of filamentary molecular clouds. 
As we have noted in Section \ref{sec:numerical}, a numerical solution is completely determined by the choice of
three dimensionless parameters; the flux to mass ratios $\Gz$, $\Gphi$, and the concentration parameter $C$.

Although $\Rs$ can be observed with little difficulty, obtaining an accurate value for $C$ is difficult because of the 
uncertainty in the core radius $r_0$.  According to equation \ref{eq:r0scale}, the core radius depends on both the central 
density and velocity dispersion along the axis of the filament, both of which might be quite uncertain.
We can, however, estimate a rough upper bound to $C$ using the data of table \ref{tab:data}.
We do not presently know whether the central (axial) velocity dispersions of filamentary clouds are 
dominated by non-thermal motions, as the bulk of the cloud certainly is, or if the velocity dispersions are thermal,
as they are in many low-mass cloud cores.  Nevertheless, we do know that that $\sigma_c$ must be at least
the thermal value, which is $0.23~km~s^{-1}$, assuming a temperature of $15 K$.  Central densities are probably 
less than about $10^4~cm^{-3}$, which is typical of a core.  Therefore, equation \ref{eq:r0scale} implies that 
$r_0$ is probably not less than $\approx 0.04~pc$.  In Table \ref{tab:data}, we find that $\Rs\appleq 0.5~pc$ for 
most (but not all) of the filaments in our sample.
Therefore, equation \ref{eq:Cdef} implies that most filamentary clouds should have concentration parameters that
are less than approximately 1.1.  
This estimate should be treated with caution, considering the uncertainties and generalizations in our calculation.  In particular, we
note that larger filaments, such as the Northern and Southern Filaments in the Orion region (See Table \ref{tab:data}) have
radii that are many times larger than the value that we used in our calculation and may, therefore, have concentration 
parameters that exceed our upper bound.  

Three observable quantities
shall be required to constrain our theoretical models.  We have previously (Section \ref{sec:obs}) found
the virial parameters $\Ps/\Pave$ and $m/\mvir$ to be useful in showing that toroidally dominated helical fields 
play an important role in the virial equilibrium of filamentary clouds.
We use these parameters, as well as a third parameter specifying the ratio of average magnetic to kinetic energy densities
to constrain our
models.  Accordingly, we define a virial parameter
\be
X=\frac{M}{K},
\label{eq:Xdef}
\ee
where $M$ and $K$ are the average magnetic and 
kinetic energy densities within the cloud defined by
\bea
M &=& \frac{\int_V (\Bz^2+\Bphi^2) dV}{8\pi V} \nn\\
K &=& \frac{3}{2} \rhoave \sigsqave,
\label{eq:energies}
\eea
and $V$ is the volume of the cloud (not to be confused with $\V$).
Myers and Goodman (1988a,b) have provided
considerable observational evidence that the average magnetic and kinetic energy densities
are in approximate equipartition, with 
$M \approx K$ to within a factor of order 2.
Therefore, we impose the auxiliary constraint that 
\be
X \approx {\cal O}[1]
\label{eq:X}
\ee
for filamentary clouds with realistic magnetic fields.
This equipartition of energy has been explained by attributing the non-thermal motions within molecular 
clouds to internally generated \Alfvenic turbulence (BM92).
Since super-\Alfvenic turbulence is highly
dissipative, the \Alfven speed poses a natural limit for the non-thermal velocity dispersion (BM92).
Thus, we expect $\sigma \approx v_A$ for molecular clouds.  Defining the average squared \Alfven speed as
\be
\langle v_A^2 \rangle = \frac{\int_0^m v_A^2 dm'}{m},
\ee
it is easy to show that
\be
X=\frac{\langle v_A^2 \rangle}{3\sigsqave}.
\ee
Therefore, $X\approx 1$ is a natural result for magnetized clouds supported against gravity by
\Alfvenic turbulence.  In the analysis that follows, we assume that
\be
0.2\leq X\leq 5
\label{eq:Xconstraint}
\ee
for all reasonable models and that $0.5\leq X\leq 2$ is appropriate for
our most realistic models.

% MONTE CARLO!!!
\subsubsection{Monte Carlo Exploration of the Parameter Space}
\label{sec:MC}
\begin{figure}

\hspace{\fill}
\begin{minipage}{\linewidth}
\psfig{file=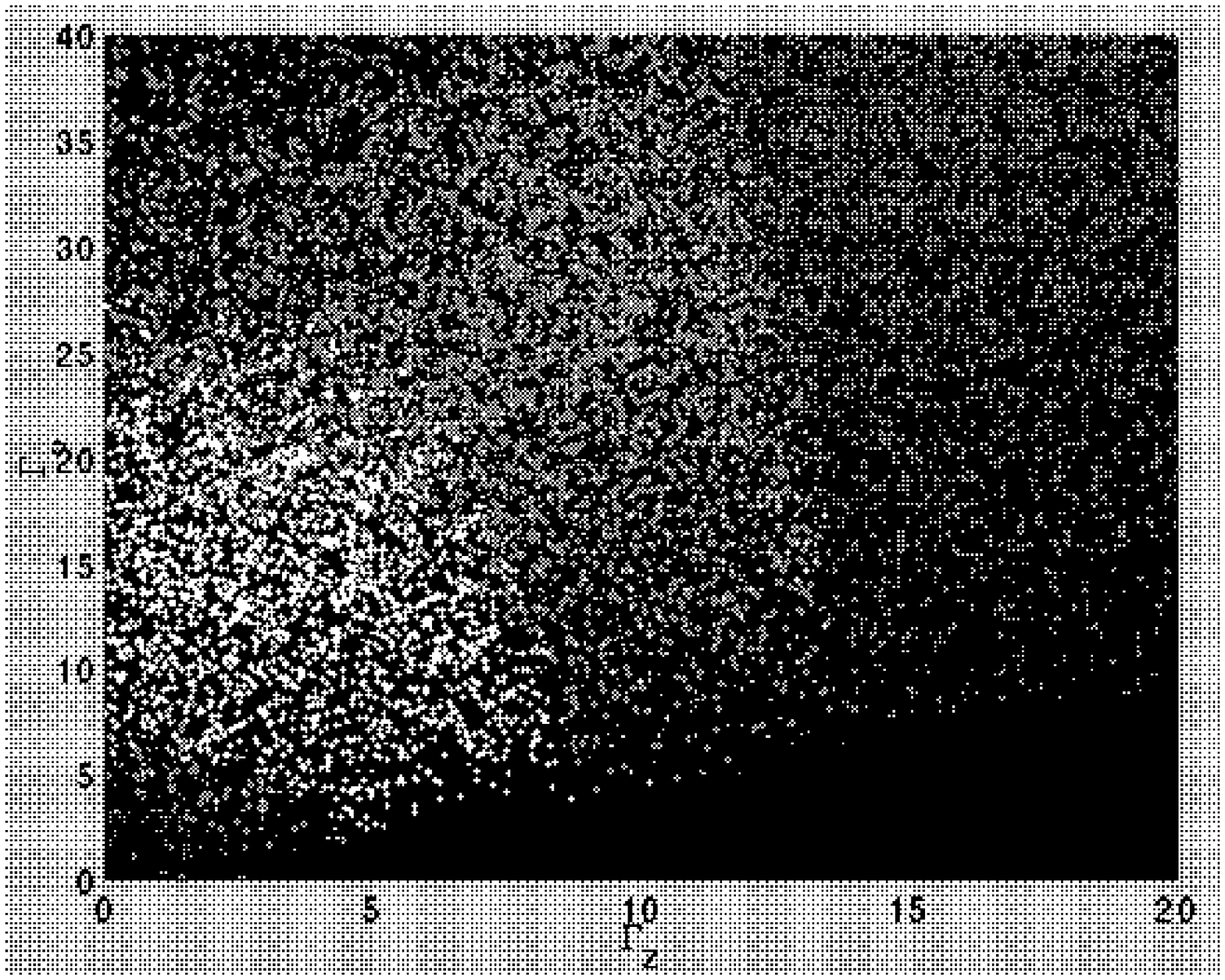,width=.87\linewidth}
\end{minipage}
\hspace{\fill}

\hspace{\fill}
\begin{minipage}{\linewidth}
\psfig{file=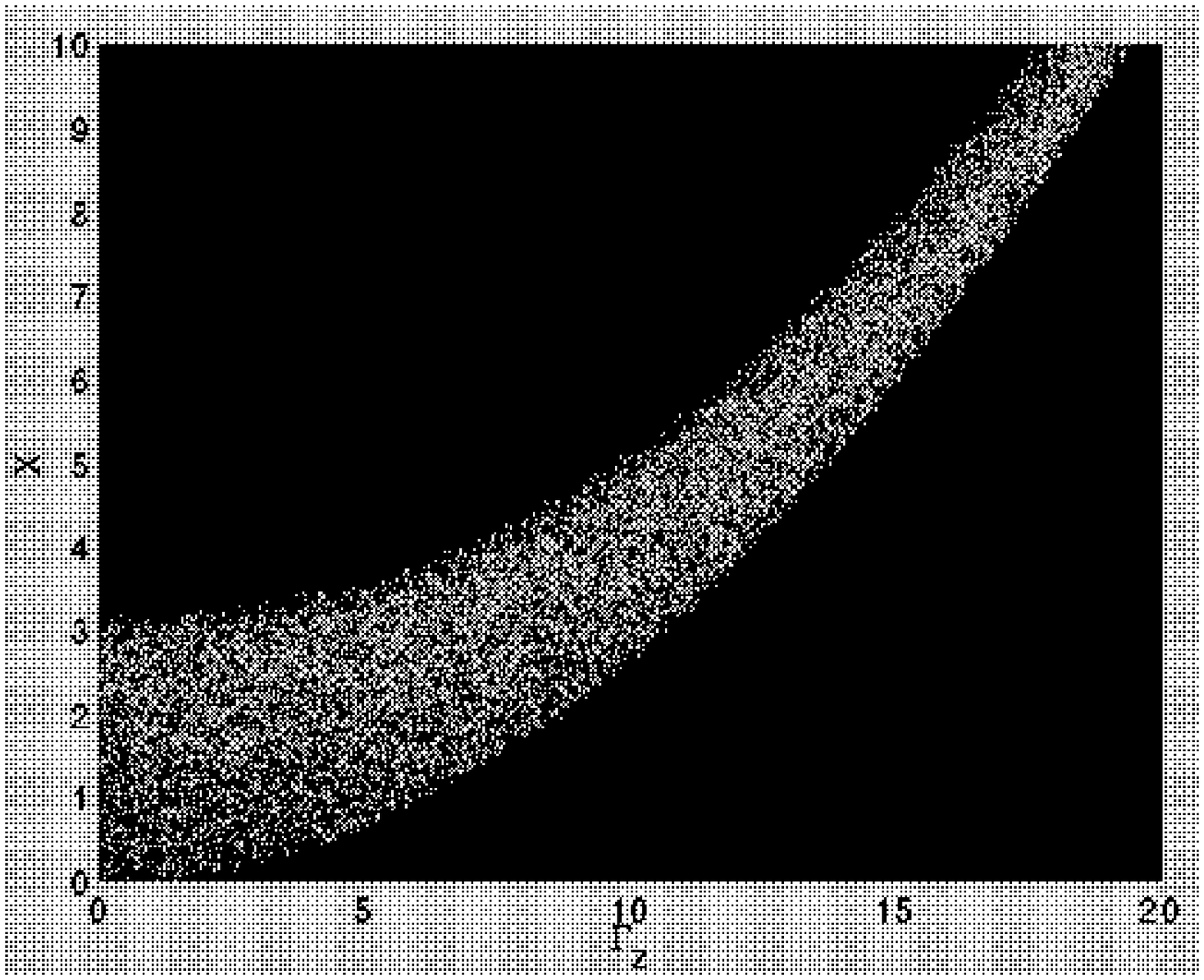,width=.87\linewidth}
\end{minipage}
\hspace{\fill}

\hspace{\fill}
\begin{minipage}{\linewidth}
\psfig{file=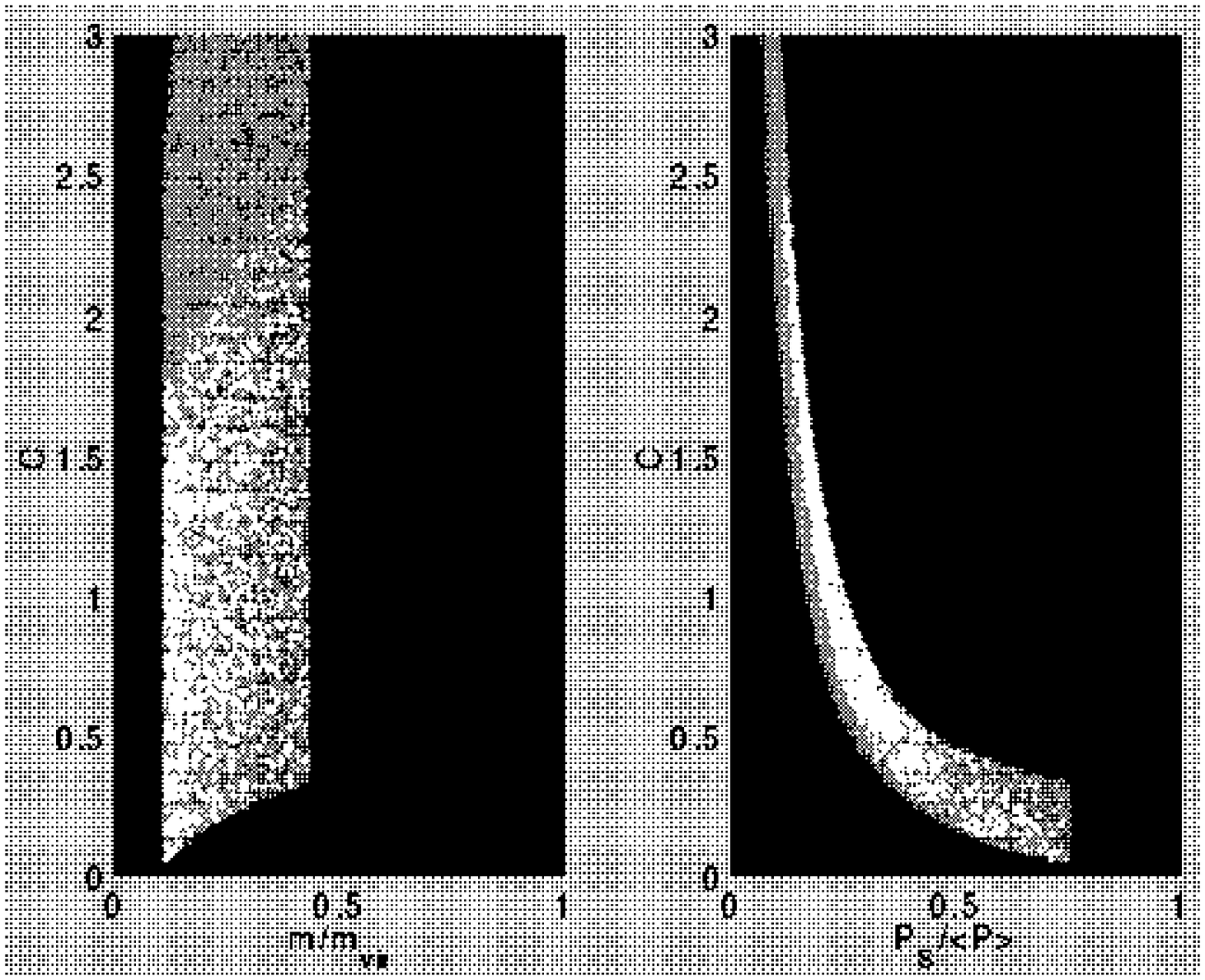,width=.87\linewidth}
\end{minipage}
\hspace{\fill}

\caption{Isothermal Models.  We show the results of our Monte Carlo exploration for isothermal filaments.  Each point on these figures
represents a model that obeys the observational constraints given in equation \ref{eq:constraints}; thus, we determine which ranges of
$\Gz$, $\Gphi$, and $C$ result in models that agree with the available observational data.  a) (top) The grayscale represents different 
ranges for $X$, as defined in equation \ref{eq:Xdef}.  The most realistic solutions, with $0.5\appleq X\appleq 2$ are shown as white dots.  The
medium gray dots have $0.2\appleq X\appleq 5$.  The dark gray dots represent models that 
are outside of these ranges with unrealistic magnetic field strengths.  b) (middle) We show that $X$ is determined mainly by $\Gz$.  
c) (bottom) We show the allowed ranges of the concentration parameter $C$.  The shading is the same as in a).  However, we note that the darkest gray dots are
mostly hidden ``behind'' the medium gray, in this case.}
\label{fig:MCiso}
\end{figure}

\begin{figure}
\begin{minipage}{\linewidth}
\psfig{file=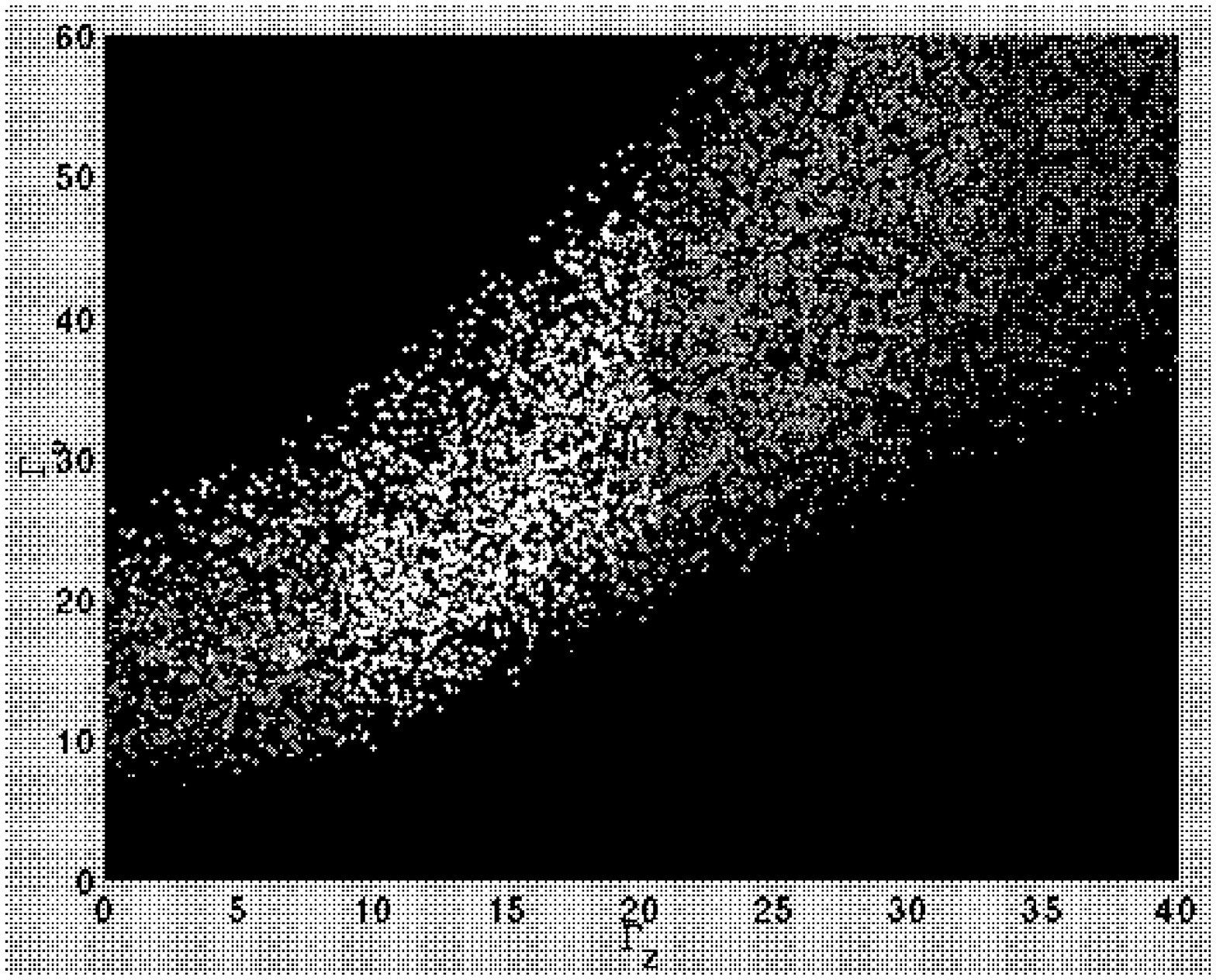,width=\linewidth}
\end{minipage}

\begin{minipage}{\linewidth}
\psfig{file=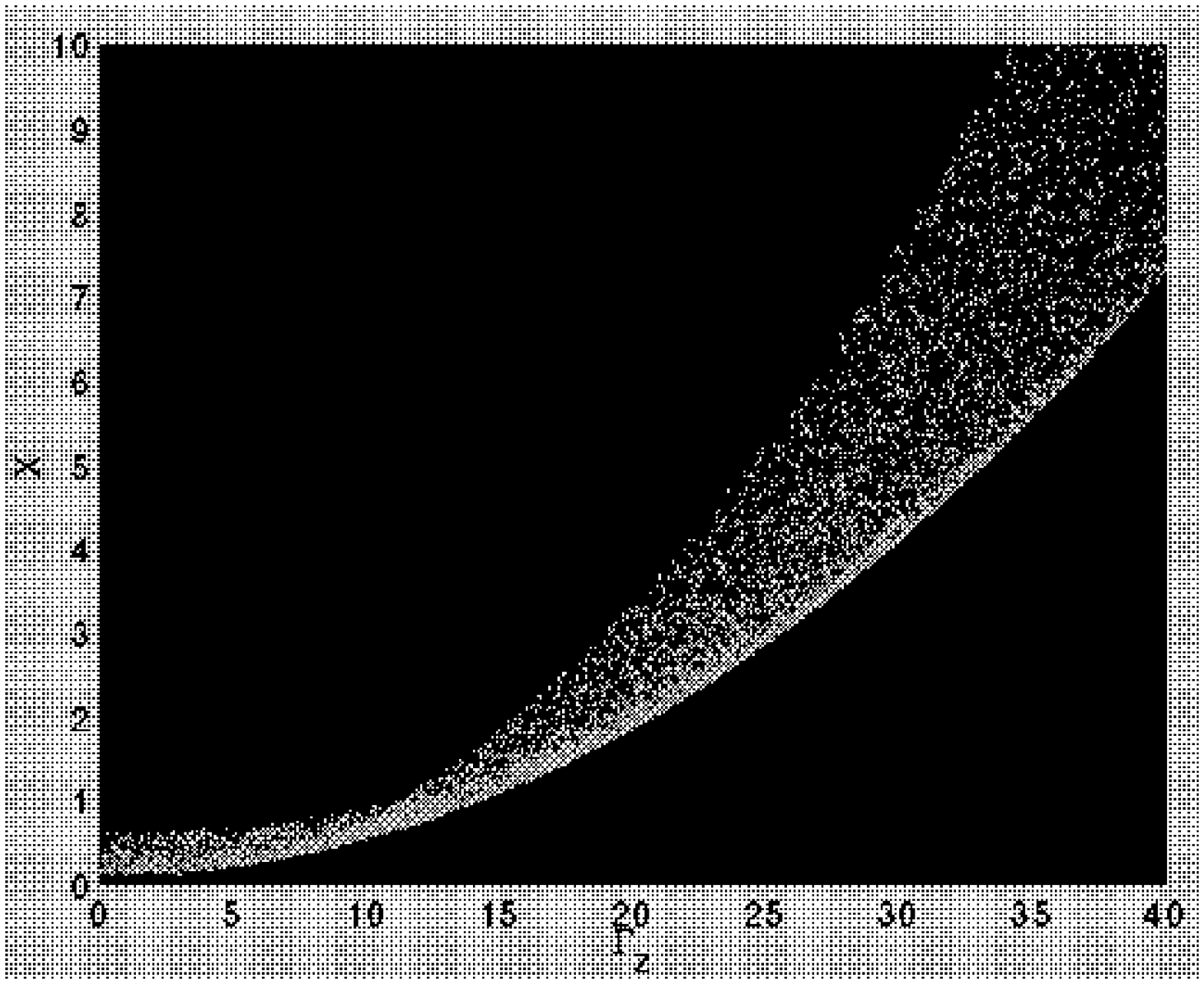,width=\linewidth}
\end{minipage}

\begin{minipage}{\linewidth}
\psfig{file=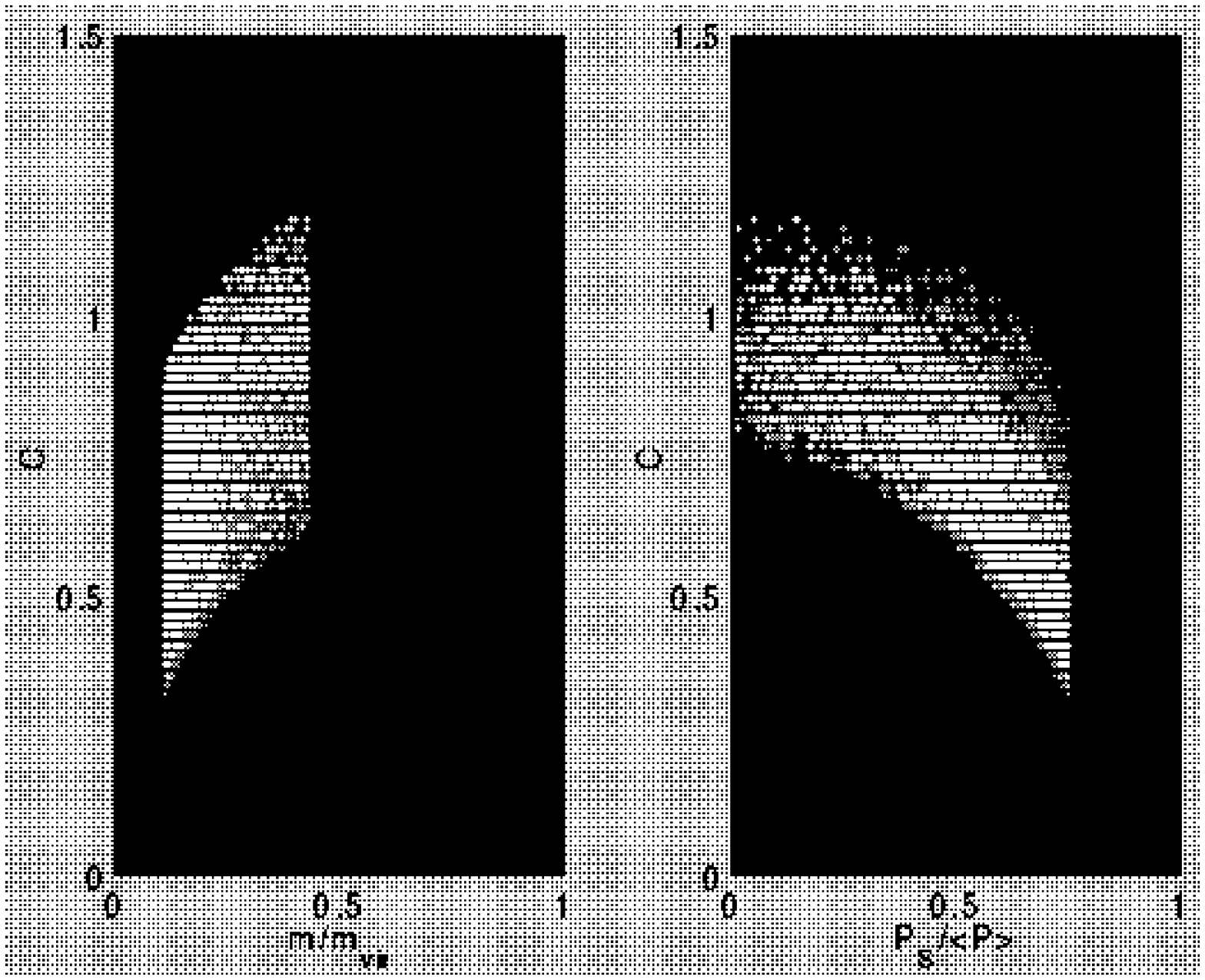,width=\linewidth}
\end{minipage}

\caption{Logatropic Models.  We show the results of our Monte Carlo analysis for logatropic filaments.
The graphs are as described in the caption of Figure \ref{fig:MCiso}.}
\label{fig:MClog}
\end{figure}

In this Section, we perform a Monte Carlo sampling of our parameter space in order to determine which values of 
$\Gz$, $\Gphi$, and $C$ result in models that obey all of our constraints.
The Monte Carlo analysis is very straightforward.  We simply assign random values to the three theoretical parameters 
and compute helical field models using the mathematical framework of Section \ref{sec:formalism}.  Once a solution has been
obtained, we compute $m/\mvir$, $\Ps/\Pave$, and $X$ using equations \ref{eq:mvir}, \ref{eq:averages}, and \ref{eq:Xdef},
from which we easily determine whether or not the solution obeys our constraints.  Figure \ref{fig:MCiso}
shows the results of our Monte Carlo analysis for isothermal models, while Figure \ref{fig:MClog} 
shows the results for logatropic models.  Each point in these figures represents a model that obeys our constraints on $m/\mvir$ and $\Ps/\Pave$;
models that fall outside of these constraints have been discarded.  

The grayscale in Figures \ref{fig:MCiso}a and \ref{fig:MClog}a
represent different ranges for $X$.  The most likely range of $X$, with $0.5\le X\le 2$ is shown as the lightest coloured points.
The next darkest gray dots represent a less likely, but still possibly allowed range, with $0.2\le X\le 5$, while the darkest gray dots represent models
that are outside of these ranges, and therefore have unrealistically large or small magnetic fields.
It should be noted that there relatively little overlap between these regions; they map out quite distinct regions on the diagrams.
From Figure \ref{fig:MCiso}a, we find that the allowed ranges for the flux to mass ratios are approximately
\bea
5 ~\appleq~ \Gphi &\appleq& 25 \nn\\
\Gz &\appleq& 8 
\label{eq:MCiso}
\eea
for isothermal filaments with $0.5\le X\le 2$.  A somewhat larger region of the parameter space is allowed
for filaments with $0.2\le X\le 5$.
Comparing with Figure \ref{fig:MClog}a, we find that the allowed flux to mass ratios are
somewhat larger for logatropic filaments, where we find
\bea
10 ~\appleq~ \Gphi &\appleq& 40 \nn\\
10 ~\appleq~ \Gz &\appleq& 20
\label{eq:MClog}
\eea
when $0.5\le X\le 2$.
We note that $\Gz$ is more tightly constrained than $\Gphi$ for both isothermal and logatropic filaments.

In Figures \ref{fig:MCiso}b and \ref{fig:MClog}b, we plot the our magnetic parameter $X$, for the allowed models, against the poloidal
flux to mass ratio $\Gz$.  We find that $X$ has a very strong dependance on $\Gz$ for both isothermal and logatropic models.  
Moreover, we find that there is no obvious correlation between $X$ and $\Gphi$.
Since we can regard $X$ as nearly a function of $\Gz$ alone, the auxiliary constraint on $X$ directly constrains
$\Gz$.  It is for this reason that somewhat tighter contraints are obtained on $\Gz$ in equations \ref{eq:MCiso} and 
\ref{eq:MClog}, than on $\Gphi$.

Figures \ref{fig:MCiso}c and \ref{fig:MClog}c show the dependence of the concentration parameter $C$ on
$m/\mvir$ and $\Ps/\Pave$ for models that are allowed by the observations.  We find that $C$ may range
from 0 to $\approx 3$ for isothermal models, but $C\appleq 1.7$ for most solutions
where $0.5\le X\le 2$.  Moreover, we find that $C$ correlates rather well with $\Ps/\Pave$, with greater
values of $\Ps/\Pave$ corresponding to smaller values of $C$.  
We note that most filamentary clouds probably have $C\appleq 1.1$,
considering our discussion in Section \ref{sec:helix}.  However,  we do not enforce this upper
bound as a rigid constraint, since further data on the central densities and velocity dispersions of filamentary clouds
needs to be obtained in order to make our argument definitive.  
We find that $C\appgeq 1$ whenever 
$\Ps/\Pave\appleq 0.25$; therefore, isothermal filaments with $C\appleq 1$ must be subject to external
pressures that are at least one fourth of the mean internal pressure.  Such relatively high external pressures
are well within the range of pressures allowed by equation \ref{eq:constraints}.
The concentration parameter $C$ is much more restricted for logatropic models, where $C$ may
range only from aprroximately 0.4 to 1.2.  
As a general trend, we find that
$C$ increases slightly with $m/\mvir$, and also as $\Ps/\Pave$ decreases.  This is a natural result, since filaments 
become more radially extended, with greater $C$, as they become closer to their critical configurations with vanishing
$\Ps/\Pave$ and maximum  $m/\mvir$.

\subsection{``Best-Fitting'' Models For Magnetized Filamentary Clouds}
\label{sec:best}

\begin{figure}
\begin{minipage}{\linewidth}
\psfig{file=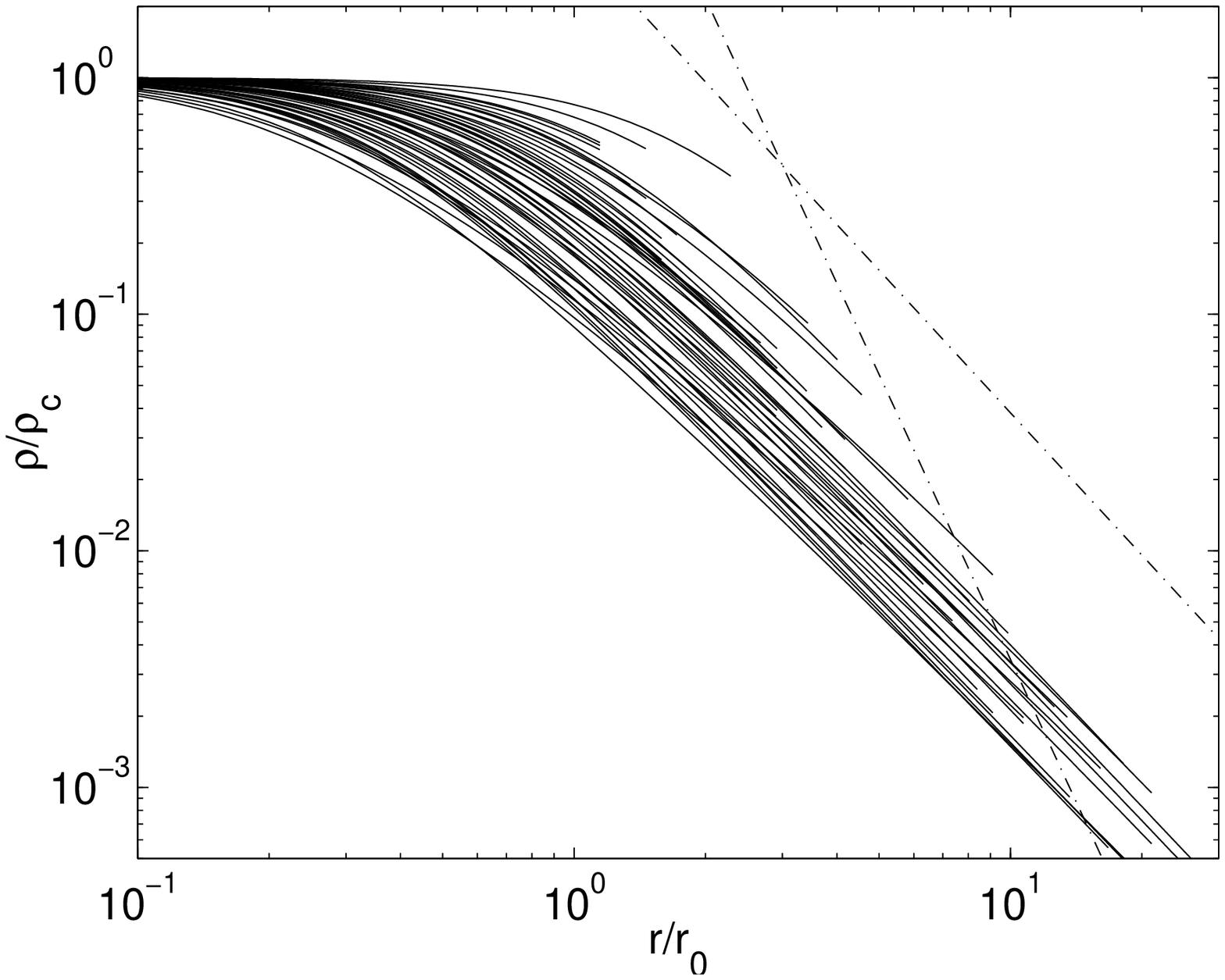,width=.9\linewidth}
\end{minipage}

\begin{minipage}{\linewidth}
\psfig{file=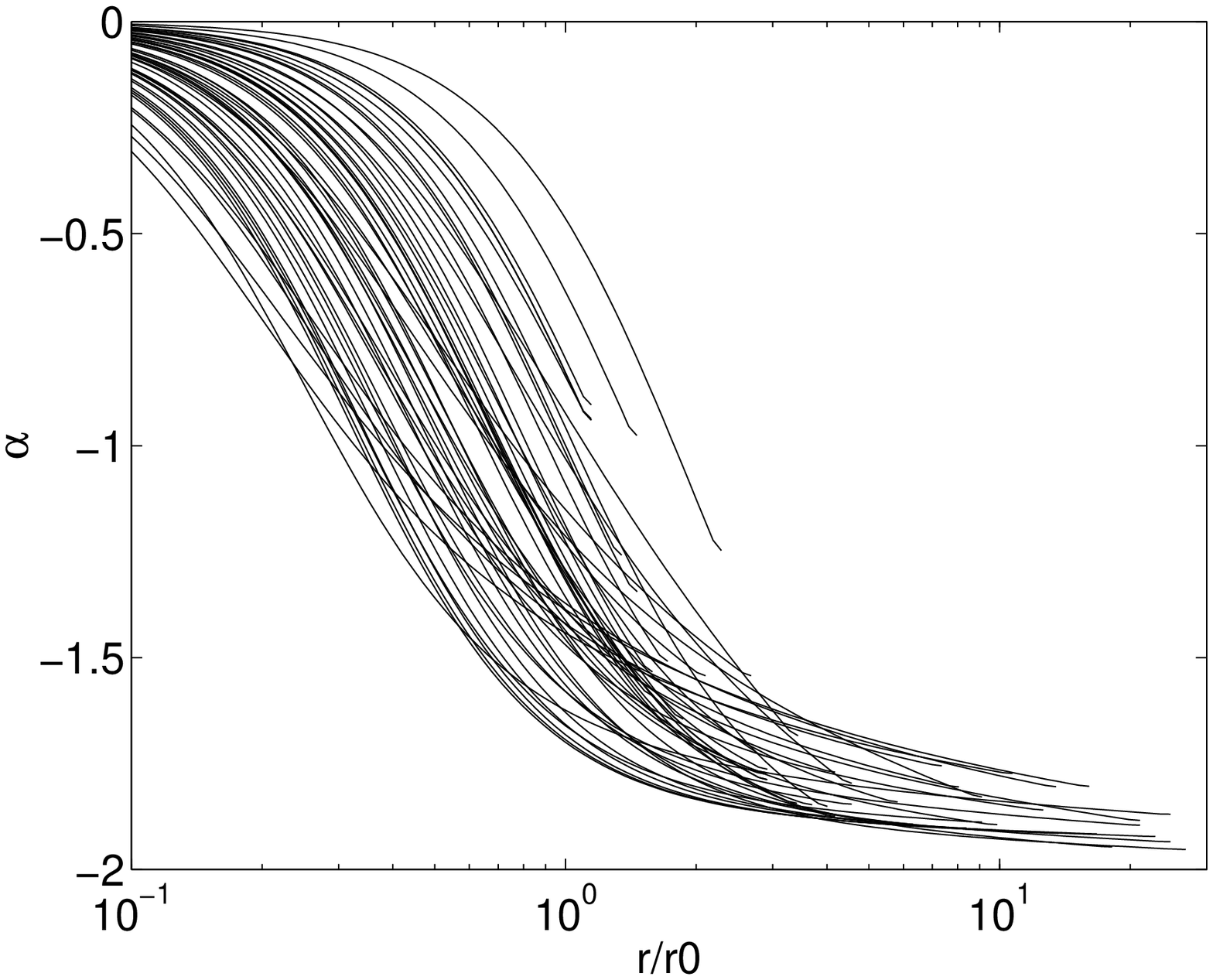,width=.9\linewidth}
\end{minipage}

\begin{minipage}{\linewidth}
\psfig{file=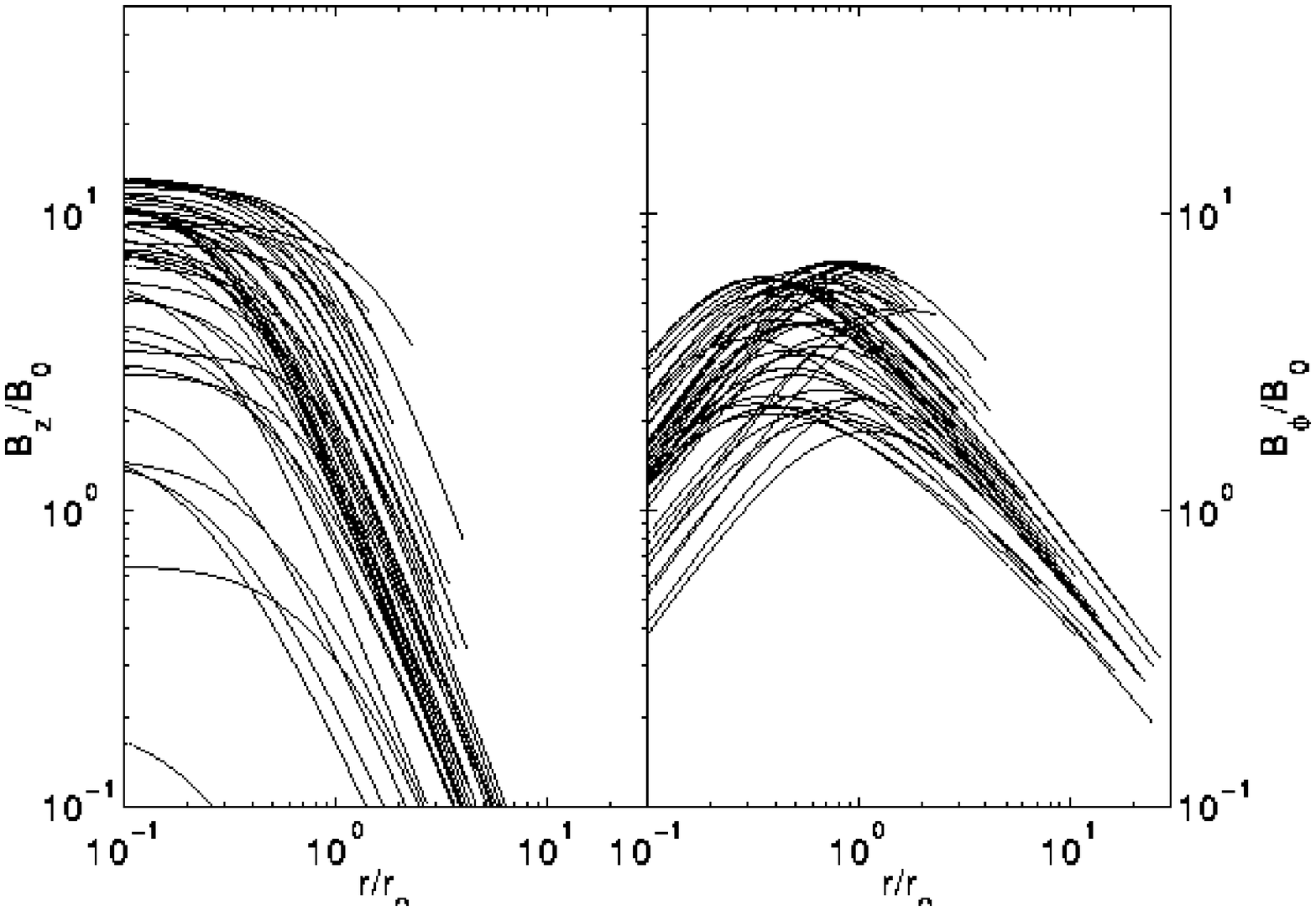,width=\linewidth}
\end{minipage}
\caption{We show a sample of 50 isothermal models that span the range of allowed parameters given in equations \ref{eq:constraints}
and \ref{eq:Xconstraint}.  a) (top) We show the density profiles of the models.  The dashed lines represent the $r^{-4}$ density profile of the
Ostriker solution and an $r^{-2}$ profile, which is in agreement with the observed density profiles of filamentary clouds (Alves et al. (1998), Lada, Alves, and Lada (1998)).
b) (middle) We show the how the power law index $d\ln{\rho}/d\ln{r}$ behaves with radius.  c) (bottom) We show the behaviour of the poloidal and toroidal
components of the magnetic field.} 
\label{fig:BESTiso}
\end{figure}

In Figure \ref{fig:BESTiso}, we show 50 isothermal helical field models that span the range of parameters allowed by 
equations \ref{eq:constraints} and \ref{eq:Xconstraint}.  We see that our allowed models possess a number of very
robust characteristics.  Most importantly, we find that most of our isothermal models have outer density 
profiles that fall off as $\sim r^{-1.8}$ to $r^{-2}$, with some of most truncated models having somewhat 
more shallow profiles.  This is most clearly shown in Figure \ref{fig:BESTiso}b, where we have plotted the power law
index $\alpha=d\ln{\rho}/d\ln{r}$ as a function of the dimensionless radius $r/r_0$.  We observe that $\alpha$ becomes
more negative with increasing radius, but that
none of our models ever have density profiles that are steeper than $r^{-2}$.  Thus, we
find that our isothermal helical field models have density distributions that are much more shallow than the $r^{-4}$
Ostriker solution.  This radical departure from the Ostriker solution is clearly due to the
dominance of the magnetic field over gravity in the outer regions.   
The overall effect of the helical field is to modify the density structure of the Ostriker solution 
so that a much more realistic form is obtained.  In particular, we note that A98 and LAL98 
have recently used extinction measurements of
background starlight in the near infra-red to show that two filamentary clouds, namely L977 and IC 5146, 
have $r^{-2}$ density distributions.
Our helical field models have density profiles that are essentially the same as those obtained for models with purely toroidal fields
in Section \ref{sec:pure}.  Therefore, we conclude that the outer density distribution is shaped primarily by the 
toroidal component of the field.  We note, however, that the toroidal field is in fact much weaker than the poloidal field
throughout most of a filamentary cloud.  In all cases, the basic magnetic structure is that of a poloidally dominated core 
region surrounded by a toroidally dominated envelope, where the field is relatively weak.

\begin{figure} 
\begin{minipage}{\linewidth}
\psfig{file=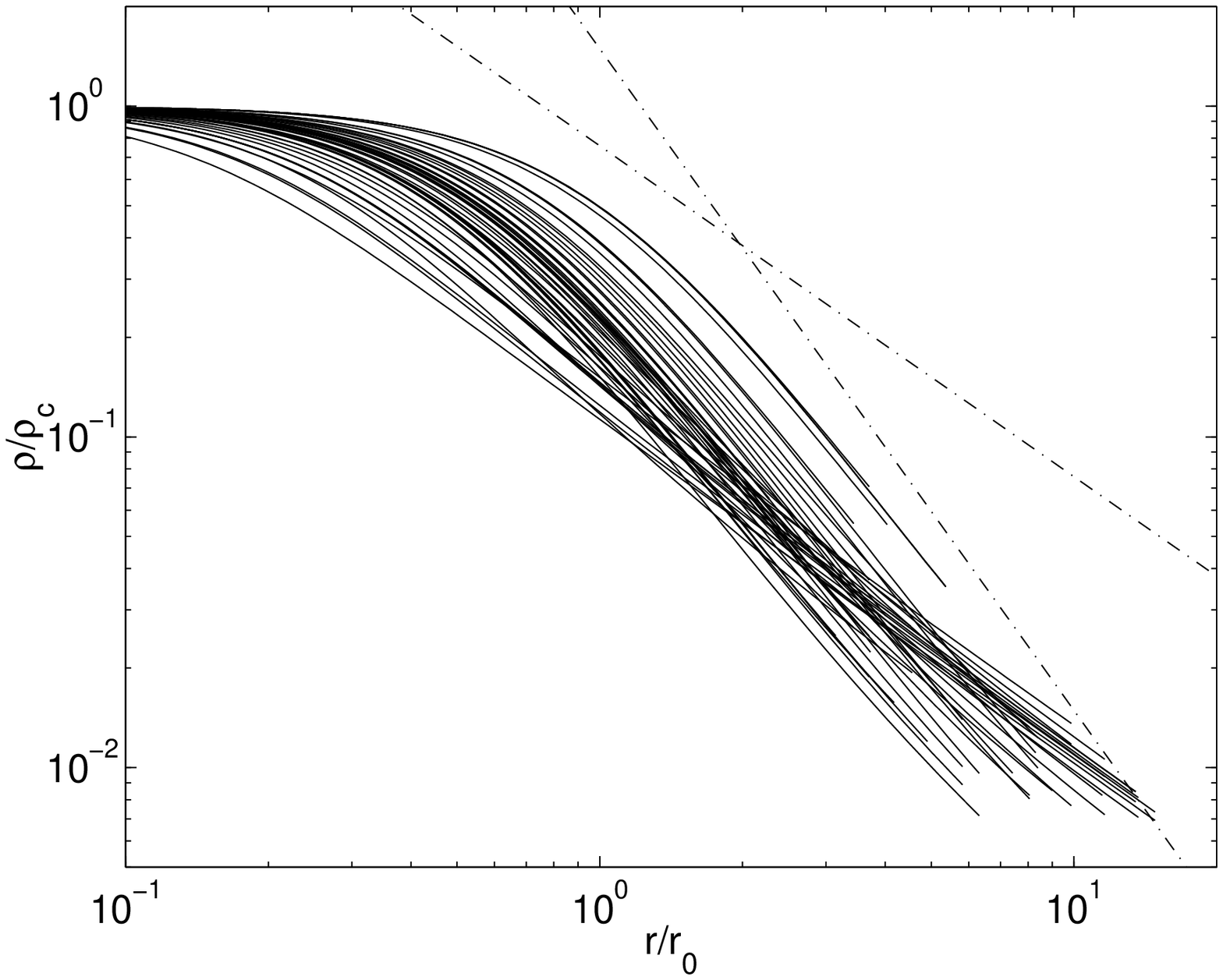,width=.9\linewidth}
\end{minipage}

\begin{minipage}{\linewidth}
\psfig{file=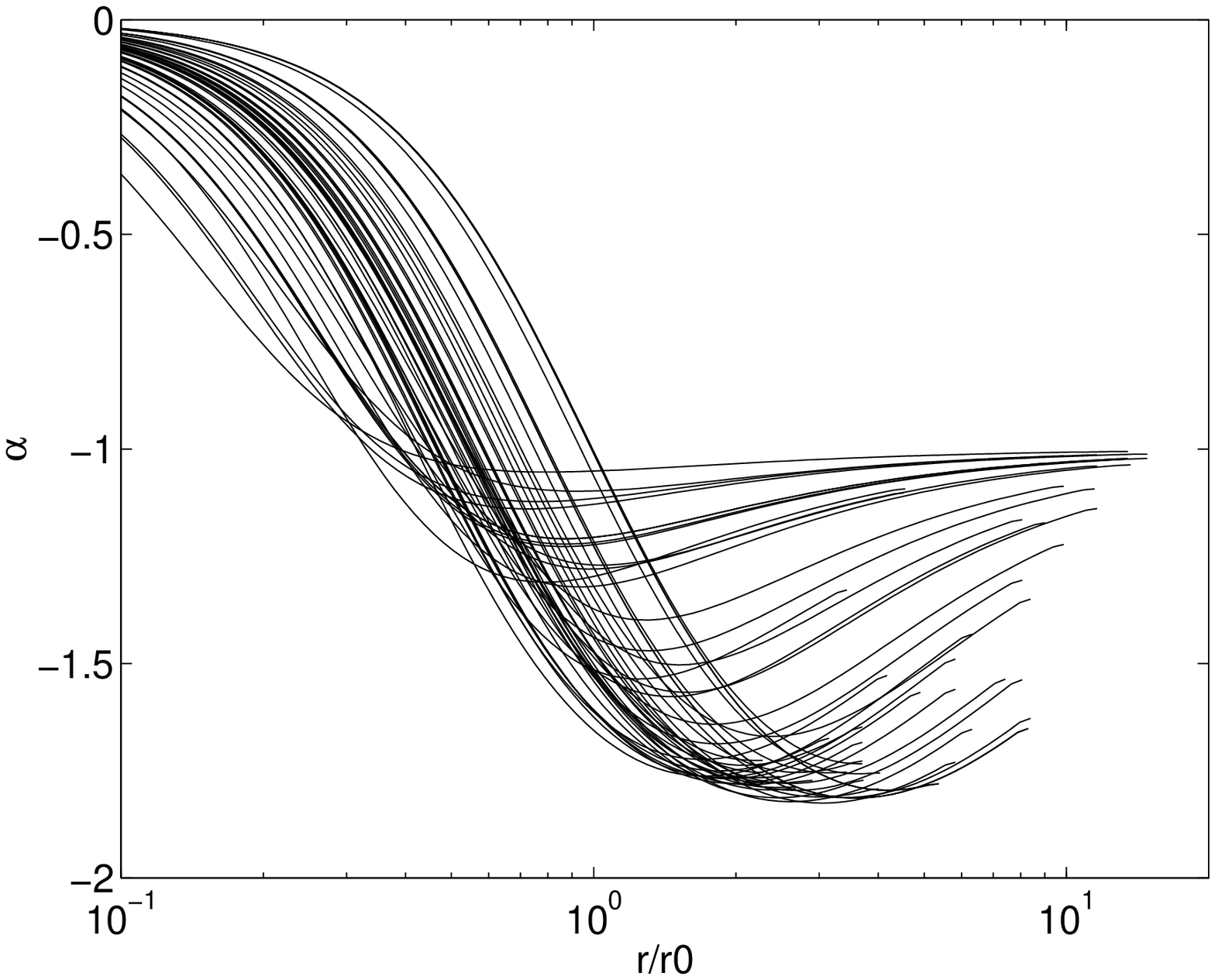,width=.9\linewidth}
\end{minipage}

\begin{minipage}{\linewidth}
\psfig{file=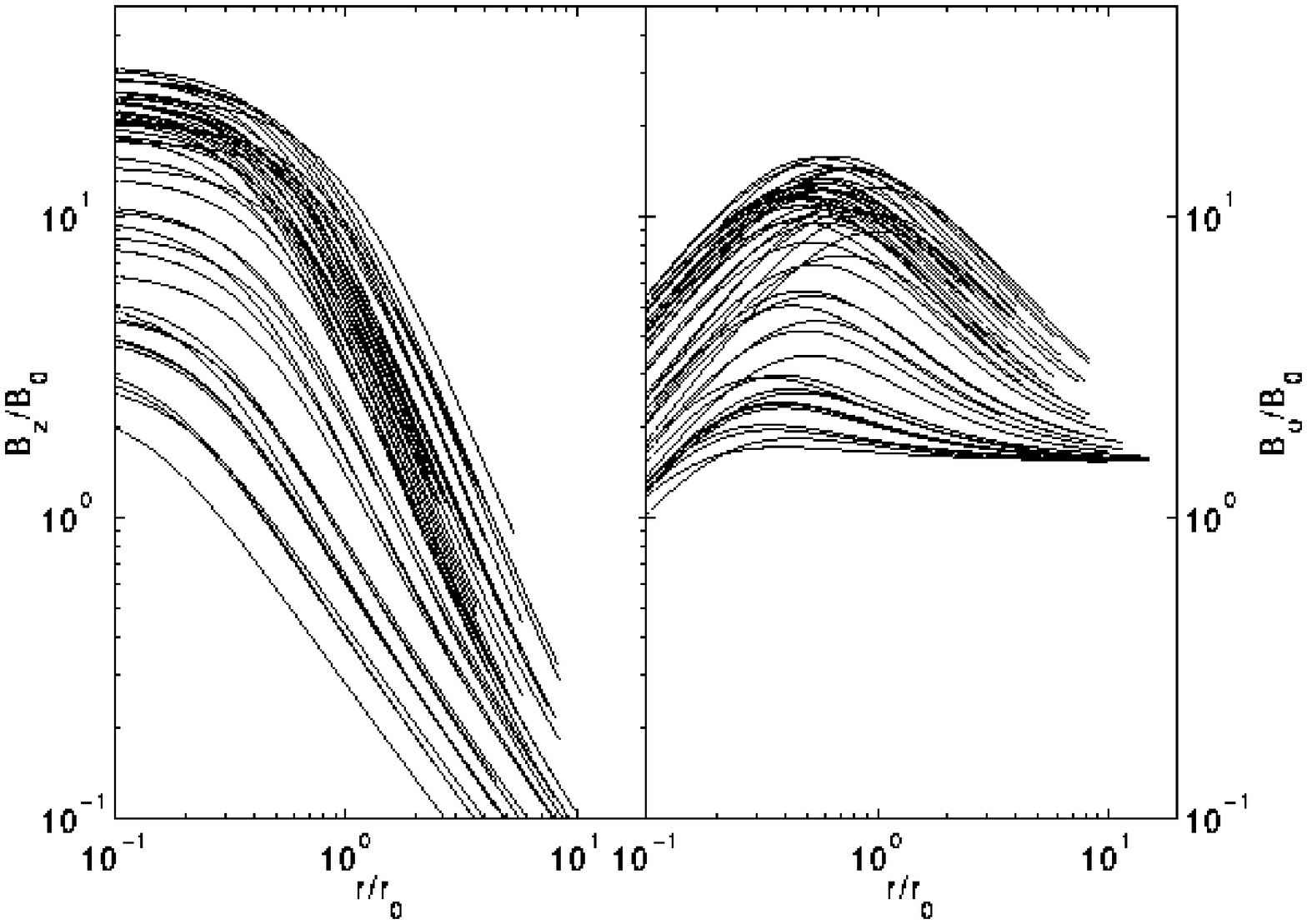,width=\linewidth}
\end{minipage}
\caption{We show a sample of 50 logatropic models that span the range of allowed parameters given in equations \ref{eq:constraints}
and \ref{eq:Xconstraint}.  a) (top) We show the density profiles of the models.  The dashed lines represent the $r^{-4}$ density profile of the
Ostriker solution and an $r^{-2}$ profile, which is in agreement with the observed density profiles of filamentary clouds (Alves et al. (1998), Lada, Alves, and Lada (1998)).
b) (middle) We show the how the power law index $d\ln{\rho}/d\ln{r}$ behaves with radius.  c) (bottom) We show the behaviour of the poloidal and toroidal
components of the magnetic field.}
\label{fig:BESTlog}
\end{figure}

In Figure \ref{fig:BESTlog}, we show a sample of 50 logatropic models that are allowed by our constraints.
The main difference between the logatropic models and the isothermal models shown in Figure \ref{fig:BESTiso} is that there 
is a much greater variety of allowed density distributions for the logatropes.  We find logatropic filaments with 
density profiles as shallow as $r^{-1}$ and as steep as $r^{-1.8}$.  Unlike the isothermal solutions, $\alpha$ does not decrease 
monatonically.  Rather, it usually reaches a minimum value somewhat less than -1
when $r/r_0\approx 1$ to $3$, and increases at larger radii.  The result is that the density distribution usually contains
a small region where the density falls quite rapidly, which is surrounded by an envelope with a more gentle power law.
Many logatropic models have density profiles that are too shallow to explain the A98 and LAL98
data.  However, we have also found many logatropic models that approach the observed $r^{-2}$ profiles.
{\em The main difference between isothermal and logatropic models is that 
isothermal filaments produce a nearly ``universal'' $r^{-1.8}$ to $r^{-2}$ density profile, while logatropic filaments 
show a much larger range of behaviour.}

\section{Discussion}
\label{sec:Discussion}

We show, in Section \ref{sec:obs}, that most of the filamentary molecular clouds in our sample have velocity dispersions
that are too high for them to be bound by gravity and surface pressure alone.
Thus, we find evidence that many filamentary clouds are probably wrapped by helical magnetic fields whose
toroidal components help to confine the gas by the hoop stress of the curved field lines.
It is important to realize that this conclusion is based on our virial analysis
of filamentary clouds (Section \ref{sec:virial}) and is, therefore, independent of either the EOS of the gas
or the mass loading of the magnetic field lines.  We find that all of the filamentary clouds 
in our sample have masses per unit length that are much lower
than the critical mass per unit length for purely hydrostatic filaments, and that filaments are quite 
far from their critical configurations, where $\Ps/\Pave\rightarrow 0$ (See equation \ref{eq:constraints}.).  
Thus, star formation in filamentary clouds
must involve fragmentation into periodic cores (cf. Chandrasekhar \& Fermi 1953), rather than overall
radial collapse.  We refer to the second paper in our series for a full analysis of this process.

We construct numerical MHD models of filamentary clouds, which we explore thoroughly in Section 
\ref{sec:MC} using Monte Carlo techniques.
We show that both isothermal and logtropic models
are consistent with the available observational data.  We find that helical fields have profound 
effects on the outer density profiles of filamentary clouds.
While hydrostatic filaments have a density profile that falls off as $r^{-4}$ (see Section \ref{sec:unmagiso}), 
nearly all of our isothermal models with helical fields have much more shallow density profiles that fall off 
as $r^{-1.8}$ to $r^{-2}$.  The toroidal component of the field is entirely responsible for these 
shalow density profiles.  Logatropic models show a a greater variety of behaviour, with
density profiles ranging from $r^{-1}$ to $r^{-1.8}$.  We note 
nearly all of our isothermal models and many of our logatropic models are in excellent agreement with 
the $r^{-2}$ density profiles of the filamentary clouds L977 and IC 5146, 
which have recently been observed by A98 and LAL98.
Helical fields, rather than the nature of the EOS, seem to explain the data by controlling
the density profiles.

The confirmation of helical fields in filamentary clouds will ultimately require direct observations
of the field structure using sub-mm polarization and possibly molecular
Zeeman observations.
Unfortunately, most polarization observations of molecular clouds have been carried out in the optical 
and near infra-red regions of the spectrum; Goodman et al. (1995) have demonstrated that such measurements are likely 
a poor indicator of magnetic field direction in cold dark clouds.  
More promising is the prospect of observing the thermal dust grain emission of dark clouds, which will hopefully become 
commonplace in the near future with instruments like the SCUBA polarimeter.  In emission, we can be assured that any
polarization is due to warm dust grains within the cloud being observed.  
The observational verification of our helical field models for filamentary 
molecular clouds will need to rely heavily on such observations.

\subsection{Observational Signatures of Helical Magnetic Fields} 
Helical magnetic fields present an interesting and unique polarization pattern.  
One might expect to find polarization vectors aligned with some average
pitch angle of the magnetic field.  However, Carlqvist (1997) has modeled the polarization pattern of helical magnetic fields
and demonstrated that this assumption is incorrect.  The signature of a helical field, assuming that the polarization percent
remains small and that the cloud is optically thin, is actually polarization vectors that are either aligned with or perpendicular
to the filament axis with a possible $90^o$ change in orientation at some radius.
The reason for this
counterintuitive behaviour is that any line of sight through the filament intersects a given radius not once, but twice.
Since the order of polarizing elements is unimportant in the limit of small polarization, the combination of
any symmetric pair of polarizing elements results in a cancelation of any oblique component of the net polarization.
Although Carlqvist's model was done for absorption polarimetry, the same reasoning holds for emission.  The overall
pattern will only reflect the dominant component of the field along any line of sight; thus, our models predict
that the innermost (poloidally dominated) regions of filamentary clouds should be dominated in emission by polarization vectors
aligned perpendicular to the filament, while the polarization should be predominantly parallel to the filament
in the (toroidally dominated) outer regions.

There is some direct evidence that filamentary clouds contain helical magnetic fields.
Optical polarization and HI Zeeman data are consistent with a helical field in the large Orion A
filament oriented at an approximately $20^o$ pitch angle relative to the axis of the filament (Bally 1989).  
While the reliability of the optical polarization data is uncertain due to the reasons discussed above,
the Zeeman observations do not suffer such ambiguity.  The line of sight component of the magnetic field has been observed
to reverse across the L1641 portion of the filament (Heiles 1987, 1989).  Such a reversal is very suggestive of
a helical field wrapped around the cloud.  More recently, Heiles (1997) has suggested that an alternate explanation 
might be that the shock front of the Eridanus superbubble has swept past the Orion A cloud causing field lines
to be stretched over the filament, thus simulating a helical pattern.  While Heiles favours this idea, he 
cannot rule out the the older idea that the field is intrinsically helical.  Regardless of the true nature of the 
$\int$-shaped filament, we observe that it is one of the few filaments shown in Figure \ref{fig:observer}
where our models do not actually require a helical field for equilibrium.

\section{Summary}
\label{sec:summary}

1. All filamentary molecular clouds are truncated by external pressure
that provides a total (thermal plus turbulent) pressure which is likely in the range of
$10^4$ to $10^5~K cm^{-3}$.

2. We have derived a new form of the virial theorem appropriate for filamentary molecular clouds
that are truncated by a realistic external pressure and contain ordered magnetic fields.
We have collected data on filamentary clouds from the literature.  We find that most of the filamentary clouds 
in our sample are constrained by 
\bea
0.11 &\appleq& m/\mvir \appleq 0.43 \nn\\
0.012 &\appleq& \Ps/\Pave \appleq 0.75. \nn\\
\eea
We use these observational constraints to show that
many filamentary clouds are likely wrapped by helical magnetic fields.

3. We have used our virial equation to derive virial relations for filaments that are 
analogous to the well-known relations for spheroidal equilibria (Chi\`eze 1987; Elmegreen 1989; MP96).
We find that the virial relations for filaments differ from the corresponding relations
for spheroids only by factors of order unity.

4. We have studied the stability of filamentary molecular clouds in the sense of Bonnor and Ebert.
We find that all filamentary molecular clouds, that are initially in
equilibrium, are stable against radial perturbations.  Thus, a cloud that is initially in equilibrium cannot be made
to undergo radial collapse by increasing the external pressure; the only way to destabilize a filament
against radial collapse is by increasing its mass per unit length beyond a critical value that depends on 
the magnetic field.  This critical value is the maximum mass per unit length for which any equilibrium is possible;
for purely hydrostatic filaments, the critical mass per unit length is given by $m_h=2 \sigsqave/G$ (See Section \ref{sec:unmagvir}).
The poloidal component of the magnetic field increases the critical mass per unit length by supporting the 
gas against self-gravity.  The toroidal field works with gravity in compressing the filament; thus, the toroidal
field decreases the critical mass per unit length.

5. There are two exact analytic solutions that can be found in the limit of vanishing magnetic field.
\Stod (1963) and Ostriker (1964) have found the equilibrium solution for unmagnetized isothermal filaments; the density  
in this solution tends to an $\sim r^{-4}$ behaviour outside of the core radius.  
A98 and LAL98 have shown that the filamentary clouds L977 and IC 5146 have density profiles that fall off
as $r^{-2}$.
If this result holds true for other filaments as well, it is unlikely that the Ostriker solution
describes real molecular filaments.  In addition, we have found a singular solution for unmagnetized logatropic
filaments.  The density profile of this filament is much more shallow; it falls off only as $r^{-1}$, which is probably
too shallow to agree with observational results.

6. We have constructed exact numerical MHD models for filamentary clouds in Sections \ref{sec:numerical} and \ref{sec:numres}.
We have considered both isothermal and logatropic equations of state, which likely bracket the true underlying EOS for filamentary
molecular clouds.   The magnetic field structure is more general than in previous studies; we have assumed only 
that the poloidal and toroidal flux to mass ratios ($\Gz$ and $\Gphi$) are constant, which we justify in Section 
\ref{sec:uniform}.  

7. Isothermal models with purely poloidal magnetic fields have density profiles 
that are even steeper than the Ostriker solution; it is unlikely that such models describe real molecular clouds.  Toroidal fields
result in density profiles that are more shallow than the Ostriker solution (typically $r^{-2}$ profiles), 
and in better agreement with observations.

8. We have performed a Monte Carlo analysis of our models, in which we randomly sample our parameter space
and then determine whether or not the resulting model agrees with the observational constraints (equations \ref{eq:constraints}
and \ref{eq:Xconstraint}).  We find both isothermal and logatropic filaments that are allowed by the data.
We find that
\bea
5 ~\appleq~ \Gphi &\appleq& 25 \nn\\
\Gz &\appleq& 8 
\eea
for isothermal filaments, and
\bea
10 ~\appleq~ \Gphi &\appleq& 40 \nn\\
10 ~\appleq~ \Gz &\appleq& 20
\eea
for logatropic filaments.

9. Our best-fitting isothermal models have density profiles that fall off as only $\sim r^{-1.8}$ to $\sim r^{-2}$,
in contrast to the $r^{-4}$ behaviour of the Ostriker solution.  These shallow profiles are entirely due
to the effects of the toroidal component of the magnetic field.  Thus, helical magnetic fields are necessary
for reasonable models of isothermal filaments.  The logatropic filaments show a greater variet of density profiles
that range from $r^{-1}$ to $r^{-1.8}$.  Thus, some of the logatropic filaments may agree with the observed $r^{-2}$
profiles (A98 and LAL98), while others may be somewhat too shallow.

\section{Acknowledgements}
The authors wish to acknowledge Christopher McKee and Dean McLaughlin for their
many insightful comments.
J.D.F. acknowledges the financial support of McMaster University and an Ontario Graduate Scholarship.
The research grant of R.E.P. is supported by a grant from the Natural Sciences and Engineering
Research Council of Canada.

\appendix
\section{Derivation of the Virial Equation for Filamentary Molecular Clouds}
\label{sec:appendixA}

The tensor virial theorem is written in its most general form as
\bea
\frac{1}{2}\frac{d^2I_{ik}}{dt^2} &=& 2 T_{ik}+\delta_{ik} \left[\int_V P dV
+M_V \right]+W_{ik}-2 M_{ik} \nn\\
&+& \int_S x_k\left(\frac{B_i B_j}{4\pi}dS_j-\frac{B^2}{8\pi}dS_i\right)
-\int_s Px_k dS_i,
\label{eq:vir}
\eea
(Chandrasekhar 1961) where we work in Cartesian coordinates $x_i$, and
the integration is over the volume $V$ of the cloud bounded by surface $S$.
Here, $P$ is the total pressure (with both thermal and non-thermal contributions) and
$I_{ik}$, $T_{ik}$, $W_{ik}$, and $M_{ik}$ are respectively the Cartesian tensors for
moment of inertia, kinetic energy, gravitational potential, and magnetic energy.
The pressure and magnetic field both contribute bulk terms that support the cloud and surface terms
that confine the gas.  The magnetic term $M_V$ is just the bulk component of the magnetic
energy:
\be
M_V = \frac{1}{8\pi} \int B^2 dV
\ee
where $B$ is the magnitude of the magnetic field.
The tensor components are given by
\bea
I_{ik} &=& \int dV \rho x_i x_k \nn\\
T_{ik} &=& \frac{1}{2} \int dV \rho v_i v_k \nn\\
W_{ik} &=& -\int dV \rho x_k \frac{\di \Phi}{\di x_i} \nn\\
M_{ik} &=& \frac{1}{8\pi}\int dV B_i B_k,
\label{eq:tensordef}
\eea
where $\rho$ is density, $P$ is pressure,
$v_i$ and $B_i$ are components of velocity and magnetic field, and $\Phi$ is the gravitational
potential.

A filamentary molecular cloud 
may be idealized as a cylinder whose length greatly exceeds it 
radius. The most appropriate scalar virial equation for such an object is obtained by taking the sum of the $x$ and $y$
(11 and 22) diagonal components of the tensor virial equation \ref{eq:vir}; the $z$ component can be
ignored because we are concerned only with equilibrium in the radial direction.
Considering the surface $S$ to be the boundary of a filament of total volume $V$, we expand
the $i$th diagonal component of equation \ref{eq:vir} using the definitions \ref{eq:tensordef}:
\bea
\frac{1}{2}\frac{d^2I_{ii}}{dt^2} &=& \int_V P dV - \int_S P x_i dS_i - \int_V \rho x_i \frac{\di \Phi}{\di x_i} dV \nn\\
&+& \frac{1}{8\pi} \int_V B^2 dV - \frac{1}{4\pi} \int_V B_i^2 dV \nn\\
&+& \frac{1}{4\pi}\int_S (x_i B_i)(B_j dS_j) - \frac{1}{8\pi}\int_S B^2 x_i dS_i,
\label{eq:vir2b}
\eea
where summation over index $j$ is implied, but no sum is to be taken over index $i$.
We have also used the fact that the kinetic energy tensor $T_{ik}$ vanishes, since the velocity is everywhere zero.
Summing over components $11$ and $22$ and setting $\ddot{I}_{11}=\ddot{I}_{22}=0$ for a filamentary cloud in
equilibrium, we easily obtain
\bea
0 &=& 2\int_V P dV -\int_S  P {\bf r}\cdot d{\bf S} -\int_V \rho{\bf r}\cdot{\bf \nabla}\Phi dV \nn\\
&+& \frac{1}{4\pi} \int_V B^2 dV  - \frac{1}{4\pi} \int_V (B_x^2+B_y^2) dV \nn\\
&+& \frac{1}{4\pi}\int_S ({\bf r}\cdot{\bf B})({\bf B}\cdot d{\bf S}) -\frac{1}{8\pi}\int_S B^2 {\bf r}\cdot d{\bf S},
\label{eq:vir3b}
\eea
where the position vector ${\bf r}$ and surface element $d{\bf S}$ are defined by
\bea 
{\bf r} &=& x\hat{x}+y\hat{y} \nn\\
d{\bf S} &=& dS_x\hat{x}+dS_y\hat{y}.
\eea
Assuming that the magnetic field is helical, it is obvious
that ${\bf B}\cdot d{\bf S}=0$; thus, the first magnetic surface term on line 3 of equation \ref{eq:vir3} is 
identically zero.
Simplifying equation \ref{eq:vir3b} and dividing by the length, we obtain
the form of the virial equation most appropriate for filamentary equilibria:
\bea
0 &=& 2\int P d\V-2\Ps\V - \int_\V \rho r \frac{\di\Phi}{\di r} d\V  \nn\\
&+& \frac{1}{4\pi}\int B_z^2 d\V - \left(\frac{\Bzs^2+\Bphis^2}{4\pi}\right) \V.
\label{eq:virialb}
\eea
It is interesting to note that this may be rewritten as 
\be
0=\frac{4}{3} {\cal U} \left( 1-\frac{\Ps}{\Pave} \right) + \W +\M,
\ee
where ${\cal U}$ is the internal energy per unit length:
\be
{\cal U}=\frac{3}{2} \int_\V \rho \sigsq d\V.
\ee
Thus, we see that our virial equation for filamentary equilibria does in fact
differ substantially from the usual form for spheroids.

\section{Bonnor-Ebert Stability of Magnetized Filaments}
We have shown, in Section \ref{sec:virial}, that all hydrodynamic filaments,
which are initially in equilibrium, are stable in the sense of Bonnor (1956) and Ebert (1955).
In this section, we examine the Bonnor-Ebert stability of uniform magnetized filaments.

\subsection{Uniform Clouds}
\label{sec:uniformcase}
We begin by considering a radial perturbation of a uniform filament in which the mass per unit length
and the velocity dispersion are conserved.
Solving equation \ref{eq:uniformcloud} for the external pressure, we easily obtain
\be
\Ps=\frac{c_\phi}{\V}+\frac{c_z}{\V^2},
\label{eq:PS}
\ee
where
\bea
c_\phi &=& m\sigsq-m^2\left(\frac{G}{2}+\frac{\Gphi^2}{8\pi}\right) \nn\\
c_z &=& \frac{m^2 \Gz^2}{8\pi}.
\label{eq:cdef}
\eea
Differentiating with respect to $\V$
and simplifying using equations \ref{eq:PS}, \ref{eq:cdef}, \ref{eq:Gz}, and 
\ref{eq:averages}, we obtain
\be
\frac{d\Ps}{d\V}=-\frac{1}{\V}\left(\Ps+\Pmag\right) < 0
\ee
for all choices of $\V$ and $\Ps$.
We conclude that all uniform filaments in equilibrium (having $m<=\mmag$) are
stable in the sense of Bonnor and Ebert.

\subsection{Non-Uniform Clouds}
\label{sec:generalcase}
We now extend the argument to general magnetized filamentary equilibria using
the mathematical framework of Section \ref{sec:formalism} and Appendix D.

Again we consider a radial perturbartion of a filamentary cloud in which the mass per unit length
$m$ is conserved.  Following MP96, we shall also require that the central velocity dispersion
$\sigma_c$ remain unchanged.   
Referring to equations \ref{eq:scales} and \ref{eq:scalings}, we may write
\be 
{\tilde m}  = \frac{m}{m_0} = \frac{\fpg m}{\sigma_c^2}.
\ee
Thus, we see that the dimensionless mass per unit length ${\tilde m}$ is also conserved during the
perturbation.  Since ${\tilde m}$ is implicitly a function of ${\tilde r}$ alone, the dimensionless
radius ${\tilde \Rs}$ must remain fixed during the perturbation.  Therefore, we find that the radial perturbation
takes the form
of a simple rescaling of the dimensionless solution with none of the dimensionless variables
perturbed whatsoever.  With this result in hand, we write
\bea
\Ps &=& P_c {\tilde \Ps} = \sigma_c^2 \rho_c {\tilde \Ps} \nn\\
\Rs &=& \sqrt{ \frac{\sigma_c^2}{4\pi G} } \rho_c^{-1/2} {\tilde \Rs},
\eea
where $\sigma_c$, ${\tilde \Ps}$, and ${\tilde \Rs}$ all remain fixed during the perturbation.
Eliminating $\rho_c$, we obtain the result
\be
\Rs=\sqrt{\frac{\sigma_c^4}{4\pi G}} ({\tilde \Ps}{\tilde \Rs}^2)^{1/2} \Ps^{-1/2}.
\ee
Since $\Rs \propto \Ps^{-1/2}$, we find that
\be
\frac{d\Rs}{d\Ps}<0
\ee
for all external pressures $\Ps$.
Therefore, we conclude that all self-gravitating filaments that are initially in a state of equilibrium
(which requires $m\le \mmag$ by equation \ref{eq:mmag}), are stable in the
sense of Bonner and Ebert.

\section{Virial Relations and Larson's Laws for Filaments} 
\label{sec:scalings}
We derive the virial relations for filamentary
molecular clouds analogous to the well known relations for spheroidal
clouds (Chi\`eze 1987; Elmegreen 1989; MP96).
Using equation \ref{eq:W} in our virial equation for filamentary
clouds (equation \ref{eq:virial}), we write
\be
0=2m\sigsqave-2\Ps \V-a m^2 G + \M
\label{eq:vir3}
\ee
We have shown in Section \ref{sec:virial} that the constant $a$ is
exactly unity for any filament but we retain the constant in order to
more directly compare with the corresponding
expressions for spheroidal clouds.
Equation \ref{eq:vir3} can be rewritten as
\be
\amag=a\frac{1-\M/|\W|}{1-\Ps/\Pave} \equiv \frac{\mvir}{m}
\label{eq:amag}
\ee
where we have introduced the observable virial parameter $\amag$
for filamentary clouds
analogous to that of Bertoldi and McKee (1992, hereafter BM92).  We may also write
\be
\amag=\anon \left( 1-\frac{\M}{|\W|} \right),
\label{eq:amag2}
\ee
where $\anon$ is just $\amag$ evaluated in the unmagnetized limit:
\be
\anon=\frac{a}{1-\Ps/\Pave}.
\label{eq:anon}
\ee

Equations \ref{eq:amag} and \ref{eq:anon} are easily solved for the
mass per unit length $m$, radius $\Rs$, average density $\rhoave$, and
surface density $\Sigma$.  In order to
compare with the corresponding results for spheroidal magnetized equilibria
(See MP96),
we present the results as general expressions with the coefficients written in
table \ref{tab:virialtable}:
\bea
m &=& C_m\frac{\sigsqave}{\amag G} \nn\\
R &=& C_R \left(\frac{\anon-a}{\anon}\right)^{1/2}
\frac{1}{\amag^{1/2}} \frac{\sigsqave}{(G\Ps)^{1/2}} \nn\\
\rhoave &=& C_\rho \left(\frac{\anon}{\anon-a}\right) \frac{\Ps}{\sigsqave} \nn\\
\Sigma &=& C_\Sigma \frac{1}{\amag^{1/2}} \left(\frac{\anon}{\anon-a}\right)^{1/2}
\left(\frac{\Ps}{G}\right)^{1/2}.
\label{eq:virialrelations}
\eea

\begin{table}
\begin{tabular}{|c|c|c|c|c|} \hline
        &  $C_m$  &  $C_R$  &  $C_\rho$  &  $C_\Sigma$  \\
\hline
filament & 2    &  $\sqrt{\frac{2}{\pi}}$ & 1 & $\sqrt{\frac{\pi}{2}} \sec{i}$ \\
spheroid & $\frac{5}{2}$ &  $5\sqrt{\frac{3}{20\pi}}$ & 1 & $\sqrt{\frac{20}{3\pi}}$ \\
\hline
\end{tabular}
\caption[The coefficients of equations \ref{eq:virialrelations}]{The coefficients of the
virial relations given in equations \ref{eq:virialrelations}.  We have included a correction
$\sec{i}$ for the projected surface density of filamentary equilibria and defined the effective mass per
unit length for a spere as $m_{sphere}=M/(2 R)$.}
\label{tab:virialtable}
\end{table}

We have included a correction for the inclination $i$ of the filament
the the plane of the sky; this only affects the expression for the
surface density.  These expressions are remarkably similar to those for
spheroidal clouds, retaining identical functional forms and
differing only by coefficients of order unity.  Of course, the mass per unit length
of a filament cannot be compared directly to the mass of a spheroid.
However, if we define an effective mass per unit length by
taking $m_{sphere}=M/(2 R)$, the resulting expression again differs from ours by only
a numerical factor of order unity.

As long as the external pressure is approximately constant, we find that
$\sigma \propto R^{1/2}$ and $\Sigma \propto const \times \sec{i}$ for filaments.  Thus, Larson's
laws apply to filamentary clouds, aside from a trivial geometric correction factor
to take into account the inclination of the filament with respect to the observer.

\section{Mathematical Framework}
\label{sec:framework}

In this Appendix, we construct the mathematical framework used to compute the numerical solutions 
described in Section \ref{sec:numerical}.
We define the magnetic pressure of the poloidal field $\Pm$ and a useful quantity $\lowbphi$ that 
depends on the toroidal field by 
\bea
\Pm &=& \Bz^2/(8\pi) \nn\\
\lowbphi &=& r^2\Bphi^2/(8\pi).
\label{eq:magpot}
\eea
Defining the effective enthalpy of the turbulent gas by
\be
dh=\frac{dP}{\rho},
\label{eq:hdef}
\ee
and introducing the magnetic potentials $\fz$ and $\fphi$
\bea
d\fz &=& \frac{d\Pm}{\rho} \nn\\
d\fphi &=& \frac{d\lowbphi}{r^2\rho},
\label{eq:f}
\eea
we may write equation \ref{eq:equilibrium} as
\be
\dr\left(h+\Phi+\fz+\fphi\right)=0.
\label{eq:equilibrium2}
\ee
We are free to specify boundary conditions for each of these potentials along the filament
axis; $h=\Phi=\fz=\fphi=0$ at $r=0$.  Thus, equation \ref{eq:equilibrium2}
can be integrated:
\be
h+\Phi+\fz+\fphi=0.
\label{eq:equilibriumint}
\ee

It is useful to define a new radial variable by the transformation
\be
s=\ln{(r/\alpha)},
\label{eq:sdef}
\ee
where $\alpha$
is a constant that is necessary only to derive
a special analytic solution in Section \ref{sec:unmagiso}. Under this transformation,
equations \ref{eq:poisson} and \ref{eq:equilibrium2} can be combined to give
\be
\frac{d^2}{ds^2} \left(h+\fz+\fphi\right)=-\Psi,
\label{eq:maineq}
\ee
where
\be
\Psi=r^2\rho=\alpha^2 e^{2s}\rho.
\label{eq:Psi}
\ee

To solve the differential equation \ref{eq:maineq}, we must
first express $h$, $\fz$, and $\fphi$ in terms of $s$ and
the new quantity $\Psi$.
Applying equation \ref{eq:hdef} to the isothermal and logatropic
equations of state and integrating, we obtain
the following formulas for the enthalpies:
\bea
h_{iso} &=& \ln\rho=\ln\Psi-2\ln\alpha-2s. \nn\\
h_{log} &=& A\left(1+\frac{1}{\rho}\right)=A\left(1+\alpha^2 e^{2s} \Psi^{-1}\right).
\label{eq:h}
\eea
We have also used the definitions of $s$ (equation \ref{eq:sdef})
and $\Psi$ (equation \ref{eq:Psi}) to write the final forms of the enthalpies in terms
of these variables.
   
Assuming constant flux to mass ratios $\Gz$ and $\Gphi$ we derive
from equations \ref{eq:Gz}, \ref{eq:Gphi}, and \ref{eq:magpot}
\bea
\Pm = \frac{\Gz^2\rho^2}{8\pi} &=& \frac{\Gz^2 \Psi^2}{8\pi \alpha^4 e^{4s}} \nn\\
\lowbphi &=& \frac{\Gphi^2\Psi^2}{8\pi}.
\label{eq:magpot2}
\eea
Substituting these relations into equations \ref{eq:f} and integrating, we obtain
\bea
\fz &=& \frac{\Gz^2}{4\pi}\left(\rho-1\right)
= \frac{\Gz^2}{4\pi}\left(\frac{\Psi}{\alpha^2 e^{2s}}-1\right) \nn\\
\fphi &=& \frac{\Gphi^2}{4\pi}\Psi,
\label{eq:f2}
\eea
where we have applied our boundary conditions that both $\fz$ and $\fphi$ vanish along the axis of the
filament where $\rho=\rho_c$ and $\Psi=0$.
We have expressed all quantities in equations \ref{eq:h} and \ref{eq:f2} in terms of $s$ and $\Psi$
alone.  Thus, equation
\ref{eq:maineq} is closed and can, at least in principle, be solved for $\Psi$.
Since $\rho$, $\Pm$, and $\lowbphi$ are written in terms of $\Psi$ (equations \ref{eq:Psi} and \ref{eq:magpot2}),
these quantities may be determined.  Finally, the poloidal and toroidal magnetic fields may be obtained from equations
\ref{eq:magpot}.

As a brief example, we show how equation \ref{eq:maineq} naturally leads to the Ostriker (1964) solution 
discussed in Section \ref{sec:unmagiso}.  From equation \ref{eq:maineq}, we easily obtain
\be
\frac{d^2}{ds^2} \left(\ln\Psi \right)=-\Psi
\ee
for isothermal equilibria.
This equation can be solved in closed form; the solution is simply
\be
\Psi=2~sech^2 s.
\ee
Converting back to $r$ and $\rho$, the equation takes the form
\be
\rho=\frac{8/\alpha^2}{(1+r^2/\alpha^2)^2}.
\ee
The boundary condition at $r=0$ is $\rho=1$ in our dimensionless units; therefore, we
require that $\alpha=\sqrt{8}$, giving the Ostriker solution of 
equation \ref{eq:rhoost}.

It is most convenient for numerical solutions to
write equation \ref{eq:maineq} as a pair of first order
equations.  This is best accomplished by writing the gravitational
acceleration as
\be
g=-\dr\Phi=-\frac{1}{r}\ds\Phi.
\label{eq:g}
\ee
Then Poisson's equation \ref{eq:poisson} becomes
\be
\frac{1}{r^2}\ds(rg)=-\rho.
\label{eq:poisson3}
\ee
Note that there is no reason for numerical solutions
to retain the constant scale factor $\alpha$
introduced in equation \ref{eq:sdef}.  For the remainder of this section, we will 
take $\alpha=1$.
With the help of equation \ref{eq:equilibriumint} and the definition of $\Psi$
(equation \ref{eq:Psi}), our numerical system becomes
\bea
\ds(h+\fz+\fphi) &=& rg \nn\\
\ds(rg) &=& -\Psi,
\label{eq:numsys1}
\eea
where $h$, $\fz$, and $\fphi$ are functions of $\Psi$ by equations \ref{eq:h} and \ref{eq:f2}.
Thus, equations \ref{eq:numsys1} 
can be rewritten as differential equations for $\Psi$ and $g$.  We write these equations explicitly below.

Denoting the derivative $\ds$ with a prime ($'$), we derive from equations \ref{eq:f}
\bea
\fz' &=& \frac{\Gz^2}{4\pi}e^{-2s} (\Psi'-2\Psi) \nn\\
\fphi' &=& \frac{\Gphi^2}{4\pi} \Psi'.
\label{eq:genf}
\eea
Using equations \ref{eq:h}, we must calculate $h'$
separately for the isothermal and logatropic equations of
state.  We note that $h'$ can be written in the general form
\be
h'=H_1(s,\Psi)+H_2(s,\Psi) \Psi',
\label{eq:genh}
\ee
where the functions $H_1$ and $H_2$ are given in table \ref{tab:htable}.

\begin{table}
\begin{tabular}{|c|c|c|} \hline
                & $H_1$                         & $H_2$                         \\
\hline
Isothermal    & -2                              & $\Psi^{-1}$                   \\
Logatropic    & $-2A e^{2s} \Psi^{-1}$          & $A e^{2s} \Psi^{-2}$          \\
\hline
\label{tab:     htable}
\end{tabular}
\caption[]{$H_1$ and $H_2$ are the functions related to the enthalpy by
equation \ref{eq:genh}. }
\label{tab:htable}
\end{table}

Using equations \ref{eq:genf} and \ref{eq:genh}, we express our system
of equations \ref{eq:numsys1} in its final form:
\bea
\Psi' &=& \frac{rg-H_1+\frac{\Gz^2}{2\pi}e^{-2s}\Psi}
{H_2+\frac{\Gphi^2}{4\pi}+\frac{\Gz^2}{4\pi}e^{-2s}} \nn\\
g' &=& -(g+e^{-s}\Psi).
\label{eq:numsys}
\eea
Since $H_2$ is positive definite, this dynamical system is regular
on the entire interval $s\in(-\infty,\infty)$.

These equations are now in a form that can be numerically integrated given appropriate
initial conditions at the axis of the filament ($r=0$).
The problem arises however that $r=0$ occurs at $s=-\infty$ in our transformed
variable; thus, we start the integration at a small but finite value of $r$.
We expect that $\rho$ tends to a constant value of unity near the axis for any non-singular
distribution.  From the definition of $\Psi$ (equation \ref{eq:Psi}), we
find that $\Psi_0 \approx e^{s_0}$ where $s_0$ is the initial value chosen
for $s$ (typically $\approx -10$).  Recalling that $g$ is the gravitational
acceleration, we apply Gauss's law to find the initial value for $g$:
\be
g_0=-\frac{r}{2}=-\frac{1}{2}e^{s_0}.
\ee

\label{lastpage}
\end{document}